\documentclass[aps,twocolumn,superscriptaddress,amsmath]{revtex4-1}

\pdfoutput = 1

\usepackage[pdftex]{graphicx,color}
\usepackage{soul}
\usepackage{amsfonts}
\usepackage{amssymb}
\usepackage{amsmath}
\usepackage{amsthm}
\usepackage{ulem}
\usepackage{enumerate}

\usepackage{subfigure}
\usepackage{rotate}
\usepackage{bm}
\usepackage{dcolumn}

\usepackage[pdftex]{hyperref}

\let\oldmarginpar\marginpar
\renewcommand\marginpar[1]{\-\oldmarginpar[\raggedleft\tiny #1]
{\raggedright\tiny #1}}

\newcommand{\avg}[1]{\left< #1 \right>}

\newcommand{\ket}[1]{|#1\rangle}
\newcommand{\braket}[2]{\langle#1|#2\rangle}

\graphicspath{{figs/}}

\begin{document}

\title{Clustering of non-ergodic eigenstates in quantum spin glasses}

\author{C. L. Baldwin}
\affiliation{Department of Physics, Boston University, Boston, MA 02215, USA}
\affiliation{Department of Physics, University of Washington, Seattle, WA 98195, USA}

\author{C. R. Laumann}
\affiliation{Department of Physics, Boston University, Boston, MA 02215, USA}

\author{A. Pal}
\affiliation{Rudolf Peierls Centre for Theoretical Physics, Oxford University, Oxford OX1 3NP, UK}

\author{A. Scardicchio}
\affiliation{Abdus Salam ICTP Trieste, Strada Costiera 11, 34151 Trieste, Italy}
\affiliation{INFN, Sezione di Trieste, Via Valerio 2, 34127 Trieste, Italy}
\date{\today}

\begin{abstract}

The two primary categories for eigenstate phases of matter at finite temperature are many-body localization (MBL) and the eigenstate thermalization hypothesis (ETH). We show that in the paradigmatic quantum $p$-spin models of spin-glass theory, eigenstates violate ETH yet are not MBL either. A mobility edge, which we locate using the forward-scattering approximation and replica techniques, separates the non-ergodic phase at small transverse field from an ergodic phase at large transverse field. The non-ergodic phase is also bounded from above in temperature, by a transition in configuration-space statistics reminiscent of the clustering transition in spin-glass theory. We show that the non-ergodic eigenstates are organized in clusters which exhibit distinct magnetization patterns, as characterized by an eigenstate variant of the Edwards-Anderson order parameter.

\end{abstract}

\maketitle

Many systems under experimental investigation as platforms for many-body localization (MBL) \cite{Altshuler1997, Basko2006, Oganesyan2007, Pal2010, Serbyn2013, Huse2014, Kjall2014, Schreiber2015, Altland2016} have long-range interactions that mediate the direct transport of excitations. This includes disordered electronic materials \cite{Ladieu1996, Ovadia2015}, ion traps \cite{Smith2016}, interacting NV centers in diamond \cite{kucsko2016thermalization, choi2016observation}, and superconducting qubit devices developed for adiabatic quantum computing \cite{Johnson2011, boixo2014evidence, boixo2016computational}. In sufficiently long-ranged systems, the proliferation of long-distance resonances precludes quantum mechanical localization \cite{Anderson1958, Logan1987, Levitov1999, Yao2014, Burin2015}, an intuitive result strongly supported by analytic work over the last half century. Nevertheless, the quantum Random Energy Model (QREM), an infinite-range spin glass, was recently shown to exhibit a phase with localized eigenstates at finite energy density \cite{Laumann2014, Baldwin2016}. The QREM provides an analytically tractable framework for studying mobility edges and configuration-space localization. This raises the obvious question of how localization survives despite the infinite-range interactions and what role it plays in more realistic long-range systems.

Some insight comes from considering the distribution of local fields --- i.e., the energy required to flip one of the system's $N$ spins relative to a given configuration. In the QREM, flipping a spin typically changes the energy by $O(N)$. Thus the quantum fluctuations which lead to the proliferation of resonances are strongly suppressed. However, short-range models have $O(1)$ local fields, and in fact, so do power-law and infinite-range systems with general $p$-body interactions. This suggests that the eigenstate-localized phase of the QREM is an exceptional case among long-range models: strict configuration-space localization cannot exist in any model with $O(1)$ local fields, since the introduction of quantum dynamics causes resonant fluctuations.

\begin{figure}[t]
\begin{center}
\includegraphics[width=1.0\columnwidth]{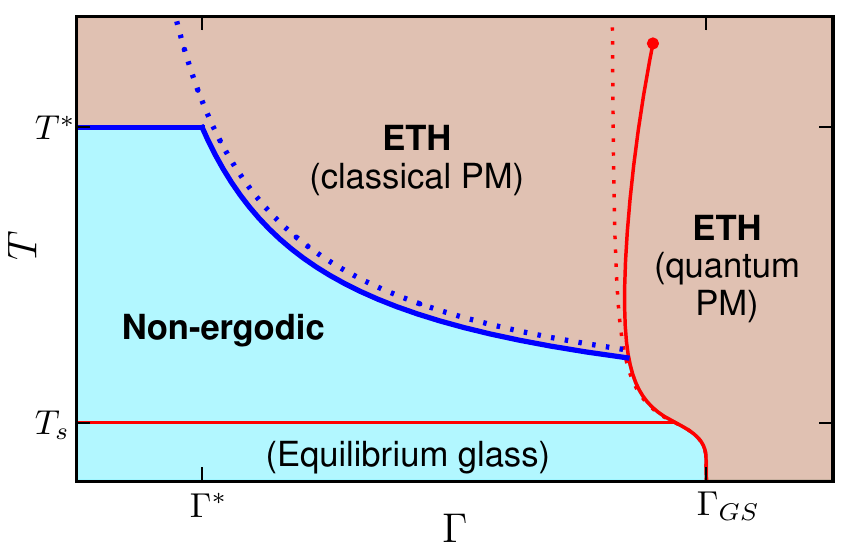}
\caption{The $T-\Gamma$ phase diagram of the quantum $p$-spin model, sketched for arbitrary $p$. The indicated $T$ and $\Gamma$ values scale as: $T_s = O(1)$, $T^* \simeq \sqrt{\frac{p}{4\ln{p}}}$, $\Gamma^* \simeq \sqrt{\frac{\ln{p}}{p}}$, $\Gamma_{GS} = O(1)$. Solid red lines are phase boundaries obtained through the imaginary-time replica formalism \cite{Goldschmidt1990a}, and the solid blue line is the eigenstate phase boundary. The corresponding dashed red and blue lines are those of the QREM ($p \rightarrow \infty$ limit).
}
\label{fig:phase_diagram}
\end{center}
\end{figure}

In this paper, we study the eigenstate properties of the quantum $p$-spin models \cite{Goldschmidt1990a, Nieuwenhuizen1998, Cugliandolo2004, Jorg2008}. Over the past four decades, these models have become paradigms for the mean-field theory of spin glasses \cite{Derrida1980, Gross1984, Gardner1985, Kirkpatrick1987}, particularly the Sherrington-Kirkpatrick model ($p = 2$) \cite{Sherrington1975, Parisi1979}. The QREM corresponds to the $p \to \infty$ limit in many senses, but the local-field distribution remains $O(1)$ in any finite-$p$ model. Consequently, the QREM's phase of localized eigenstates gives way to a phase of delocalized yet non-ergodic eigenstates (blue in Fig.~\ref{fig:phase_diagram}), similar to what was observed in the context of single-particle localization on the Bethe lattice \cite{Biroli2012, DeLuca2014, Altshuler2016}. This is in contrast to the fully-delocalized paramagnetic phase (orange in Fig.~\ref{fig:phase_diagram}), in which eigenstates satisfy the eigenstate thermalization hypothesis (ETH) \cite{Deutsch1991,Srednicki1994,Rigol2008} and exhibit thermal behavior. 

The formal distinction that we make between ergodic, non-ergodic, and MBL eigenstates concerns the off-diagonal matrix elements of local operators between them. 
Schematically, denote by $\hat{\sigma}$ any operator suppported on $O(1)$ spins and consider the state $\hat{\sigma} \ket{\Psi}$ (with $\ket{\Psi}$ an eigenstate). According to ETH \cite{Srednicki1999}, the overlap with any other eigenstate $\ket{\Phi}$ at the same energy density $\epsilon$ should scale as $\braket{\Phi}{\hat{\sigma} | \Psi} \sim \frac{1}{\sqrt{\exp{(Ns_{\textrm{eq}}(\epsilon))}}} g(\epsilon, E_{\Psi} - E_{\Phi})$, where $s_{\textrm{eq}}(\epsilon)$ is the thermodynamic entropy density and $g$ is a smooth function of $\epsilon$ and the energy difference. 
Our analysis below suggests that the eigenstates of the ETH phase in Fig.~\ref{fig:phase_diagram} obey this scaling. 
On the other hand, in an MBL phase $\hat{\sigma} \ket{\Psi}$ should have significant weight only on $O(1)$-many eigenstates \cite{Basko2006}, a notion one can make precise through a participation ratio (e.g., $\sum_{\Phi} | \braket{\Phi}{\hat{\sigma} | \Psi} |^4 \sim O(1)$). 
We find that the eigenstates of the non-ergodic phase \textit{do not} obey this definition of MBL, even though they violate ETH. Rather, they are organized into ``clusters'' (defined below). Within a cluster $c$, eigenstates follow ETH-type scaling:
\begin{equation} \label{eq:cluster_ETH}
\braket{\Phi^{(c)}}{\hat{\sigma} | \Psi^{(c)}} \sim  \frac{1}{\sqrt{\exp{(Ns_c(\epsilon))}}} g_c(\epsilon, E_{\Psi} - E_{\Phi}) , 
\end{equation}
but off-diagonal matrix elements between clusters are heavily suppressed ($c\ne c'$),
\begin{equation}
\braket{\Phi^{(c')}}{\hat{\sigma} | \Psi^{(c)}} \ll \frac{1}{\sqrt{\exp{(Ns_{\textrm{eq}}(\epsilon))}}}. 
\end{equation}
Here,  $\ket{\, \cdot \, ^{(c)}}, \ket{\, \cdot \, ^{(c')}}$ are eigenstates belonging to each cluster. $s_c(\epsilon)$ is the entropy density within $c$ (which is strictly less than $s_{\textrm{eq}}(\epsilon)$), and $g_c$ is a smooth, cluster-dependent $O(1)$ function. 
More physically, such non-ergodic eigenstates are thermal within a cluster but not thermal in configuration space as a whole.

Concretely, the quantum p-spin models are defined by
\begin{equation} 
\label{eq:p-spin_definition}
H_p = - \! \! \! \! \sum_{(i_1 \ldots i_p)} J_{i_1 \ldots i_p} \hat{\sigma}_{i_1}^z \cdots \hat{\sigma}_{i_p}^z - \Gamma \sum_{i=1}^N \hat{\sigma}_i^x \, \equiv \, H_p^C + H^Q,
\end{equation}
where the classical term $H_p^C$ sums over all distinct $p$-tuples of $N$ spins and the quantum term $H^Q$ provides a uniform transverse field.
The random couplings $J_{i_1 \ldots i_p}$ are i.i.d. Gaussians of mean 0 and variance $\frac{p!}{2N^{p-1}}$, to ensure extensivity.
It is known \cite{Goldschmidt1990a} that the thermodynamic free energy of $H_p$ approaches that of the QREM as $p$ increases. 
The eigenstate phases of $H_p$ do as well, yet the eigenstates are never localized at any finite $p$.
They are instead non-ergodic, in a manner that comes to resemble localization as $p$ increases.
We show this by studying the eigenstates within perturbation theory and the forward-scattering approximation \cite{Pietracaprina2016}.

Before we turn to detailed analysis, it is useful to consider the $p$-spin models in terms of Anderson localization on the $N$-dimensional hypercube defined by the $\sigma^z$ configuration space. $H_p^C$ is then a random potential and $H^Q$ causes hops along the edges of the hypercube.
The QREM corresponds to an uncorrelated Gaussian random potential of bandwidth $\sqrt{N}$. 
This bandwidth models that of a many-body system with extensive spectrum, but the lack of correlations implies unrealistically large local fields. 
In the $p$-spin model, the potential remains Gaussian but exhibits correlations which restrict the energy differences between adjacent sites to be $O(p)$. 
This leads to the entropically large clusters over which the eigenstates delocalize at short fractional Hamming distance (see below). 
The phase transition at finite transverse field shown in Fig.~\ref{fig:phase_diagram} corresponds to eigenstates tunneling \textit{between} clusters.
{}
To obtain the eigenstates, we must first consider the correlations in the classical energy landscape in more detail. 
At Hamming distance $Nx$ from a given configuration with energy $N \epsilon_0$, the Gaussian random potential obeys a conditional distribution \cite{Derrida1980},
\begin{equation} \label{eq:conditional_classical_energies}
P_x(\epsilon) \propto \exp{\left( -N \frac{(\epsilon - (1 - 2x)^p \epsilon_0)^2}{1 - (1 - 2x)^{2p}} \right) }.
\end{equation}
Intuitively, configurations at distance $x \lesssim \frac{1}{p}$ have similar energies and reduced fluctuations relative to independently-sampled states, as a result of the $O(1)$ local fields.
See \citep[Sec.~B]{SuppInfo2017} for more details.

Using Eq.~\eqref{eq:conditional_classical_energies}, the average number of states at fractional distance $x$ with energy density $\epsilon$ matching $\epsilon_0$ is $\binom{N}{Nx} P_x(\epsilon_0) \sim e^{Ns(x)}$, with
\begin{equation} \label{eq:annealed_position_entropy}
s(x) = -x \ln{x} - (1 - x) \ln{(1 - x)} - \frac{1 - (1 - 2x)^p}{1 + (1 - 2x)^p} \epsilon_0^2.
\end{equation}
In order to compare with the literature, it is useful to parametrize $\epsilon_0$ through the temperature $T$ defined by formal Legendre transform, even when the system fails to thermalize dynamically.
It is shown in \citep[Sec.~A]{SuppInfo2017} that $\epsilon_0 = - \frac{1}{2T} + O \left( \frac{1}{p^2} \right) $. 
Eq.~\eqref{eq:annealed_position_entropy} is an annealed average which provides a rigorous upper bound for the typical number of states, since $\mathbb{E} \left[ \ln{\ldots} \right] \leq \ln{\mathbb{E} \left[ \ldots \right] }$. 
When $s(x) < 0$, we know with certainty that there are no configurations at $x$ with energy density $\epsilon_0$.

\begin{figure}[t]
\begin{center}
\includegraphics[width=1.0\columnwidth]{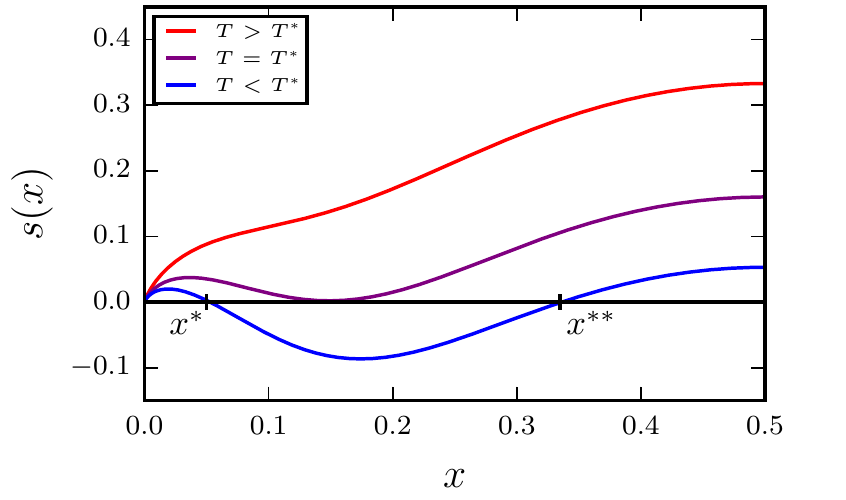}
\caption{The annealed Hamming-distance-resolved entropy $s(x)$ (for $p = 6$). $x$ is the fractional Hamming distance relative to a configuration conditioned to have energy density $\epsilon_0 = -\frac{1}{2T}$. Note that these curves are for $\Gamma = 0$. Red curve: $T = 0.83$. Purple curve: $T = 0.69$. Blue curve: $T = 0.63$.}
\label{fig:entropy_curve}
\end{center}
\end{figure}

The entropy $s(x)$  is plotted in Fig.~\ref{fig:entropy_curve} for $p = 6$ as illustration. 
A transition occurs at the temperature $T^*$. 
At $T > T^*$, there are configurations over the entire range of $x$, whereas at $T < T^*$, there is a ``forbidden'' region $(x^*(T), x^{**}(T))$ in which no configurations lie.
We have rigorously confirmed the presence of three distinct regions via a second-moment analysis, analogous to that done in Ref.~\cite{mezard2005clustering}. See \citep[Sec.~D]{SuppInfo2017} for details.
Thus the configurations at energy $\epsilon_0$ form disconnected clusters of Hamming size $x^*(T)$. Energy levels within a cluster are highly correlated, but those of different clusters are essentially independent (see Eq.~\eqref{eq:conditional_classical_energies}).
This behavior is analogous to the clustering observed in, e.g., $k$-SAT problems \cite{mezard2005clustering}, coloring of random graphs \cite{Mulet2002}, and the replica theory of spin glasses \cite{Mezard1987,Mezard2009}.

By setting $s(x) = \partial_x s(x) = 0$, we find that
\begin{equation} \label{eq:T_d_behavior}
T^* = \sqrt{\frac{p}{4\ln{p}}} \left( 1 + O \left( \frac{\ln{\ln{p}}}{\ln{p}} \right) \right) .
\end{equation}
Note that $T^* \rightarrow \infty$ as $p \rightarrow \infty$. Furthermore, at $T \ll T^*$, $x^*(T) \sim e \cdot \exp{\left( -\frac{1}{4T^2} p \right) }$ whereas $x^{**}(T) \sim O(1)$ (with respect to $p$). Thus the clusters below $T^*$ are well separated, and they become arbitrarily small as $p$ increases. Regardless, the clusters cover a macroscopic Hamming distance at any finite $p$.

With this understanding of how the \textit{classical} states at energy density $\epsilon_0$ are organized, we now introduce a small transverse field $\Gamma$ and study the eigenstates within perturbation theory. Let $\ket{\Psi_{\alpha}}$ be the eigenstate that results from perturbing the classical configuration $\ket{\alpha}$, and let $\ket{\beta}$ be another classical state separated by Hamming distance $Nx$. 
Since the perturbation $-\Gamma \sum_i \hat{\sigma}_i^x$ flips a single spin at each order, the leading non-zero contribution to $\braket{\beta}{\Psi_{\alpha}}$ arises at the $Nx$'th order. 
In the forward-scattering approximation (FSA), we retain only this contribution for each configuration $\ket{\beta}$. 
Note that many terms nonetheless contribute to $\braket{\beta}{\Psi_{\alpha}}$: one for each of the $(Nx)!$ distinct \textit{sequences} of spin-flips that transform $\ket{\alpha}$ into $\ket{\beta}$. Thus, within the FSA,
\begin{equation} \label{eq:forward_scattering_approximation}
\braket{\beta}{\Psi_{\alpha}} \approx \frac{\Gamma}{E_{\alpha} - E_{\beta}} \sum_{\mathcal{P}} \prod_{\gamma \in \mathcal{P}} \frac{\Gamma}{E_{\alpha} - E_{\gamma}}.
\end{equation}
The sum runs over the sequences $\mathcal{P}$, and the product runs over each intermediate configuration $\ket{\gamma}$ along sequence $\mathcal{P}$. 
Note that Eq.~\eqref{eq:forward_scattering_approximation} is indeed $Nx$'th order in $\Gamma$. 
See~\cite{Baldwin2016} and~\cite{Pietracaprina2016} for more explicit derivations.

% resonances at early stages should be regulated by self-energy corrections
% neglecting those corrections provides an overestimate of amplitudes at later distances
% Nonetheless, forbidden region is powerful enough to suppress tunneling at low temperatures/fields
% neglecting those corrections misses the 'random matrix-like' expansion of the state over degenerate configurations
% distinction between degenerate and resonant
Before turning to a quantitative analysis of Eq.~\eqref{eq:forward_scattering_approximation}, let us sketch the key ideas. 
Even for small $\Gamma$, the amplitude $\braket{\beta}{\Psi_{\alpha}}$ can be large if the denominators in Eq.~\eqref{eq:forward_scattering_approximation} are small. 
We will find that such ``resonances'' show up at small distances $x$ regardless of $T$ (i.e. $\epsilon_0 \equiv E_\alpha/N$) and $\Gamma$. 
The large amplitudes appear to invalidate our perturbative expansion. However, a more accurate treatment regulates them by introducing self-energy corrections \cite{Anderson1958}.
Furthermore, for $T < T^*$ there is the tunneling region $(x^*, x^{**})$ in which resonances cannot exist (see Fig.~\ref{fig:entropy_curve}). Here the self-energy corrections are negligible. Thus the na\"ive FSA accurately estimates the suppression of amplitude due to tunneling through this forbidden region.
If it predicts that \textit{every} amplitude at $x > x^{**}$ is exponentially suppressed, then we know that the eigenstates do not delocalize across the forbidden region and are nonergodic.
There turns out to be a critical $\Gamma_c(T)$ below which eigenstates are nonergodic in precisely this sense. 

Rather than introducing self-energies, an alternative approach to account for the short distance resonances is degenerate perturbation theory. 
Although precise calculations along these lines are infeasible, we expect the resulting eigenstates have amplitudes uniformly distributed across all resonant configurations, as in random matrix theory.
Restarting the perturbation theory from these hybridized states leads to new resonances which must themselves be included in the degenerate perturbation theory, leading to yet further resonances, and so on.
At $T < T^*$, this process terminates when all degenerate states at $x < x^*$ have been incorporated. We accordingly expect the eigenstates to appear thermal \textit{with respect to} this short-distance cluster (cf. Eq.~\eqref{eq:cluster_ETH}).
If $\Gamma < \Gamma_c(T)$, the eigenstates are nonetheless non-ergodic for the reasons outlined above.
Yet if $\Gamma > \Gamma_c(T)$, we find further resonances in other clusters. Since these states hybridize not just within but \textit{between} clusters, we expect them to be fully ergodic.
Similarly, at $T > T^*$ there is no forbidden region and nothing prevents every configuration at $T$ from hybridizing. Here we expect full ergodicity at any $\Gamma$. This is illustrated in Fig.~\ref{fig:phase_diagram}.

We now quantitatively demonstrate the existence of the nonergodic phase at $T < T^*$ and locate its phase boundary $\Gamma_c(T)$. 
Specifically, we count the number of resonant configurations $\beta$ at distance $x$, i.e., those that have $|\braket{\beta}{\Psi_{\alpha}}| \geq A$, where $A$ is any $O(1)$ number. 
We evaluate the sum over paths in Eq.~\eqref{eq:forward_scattering_approximation} using replica analysis \cite{Mezard1987} \citep[Secs.~E and~F]{SuppInfo2017} to find its typical behavior. 
When $|E_\alpha - E_\beta|$ is less than the resulting tunneling amplitude, $\beta$ is resonant.
The expected number of resonances at distance $x$ is then given by $e^{Nf(x)}$ \citep[Sec.~F]{SuppInfo2017}, where
\begin{widetext}
\begin{equation} \label{eq:num_resonances}
f(x) = \underbrace{\int_0^x \textrm{d} y \ln{\frac{1}{1 - (1 - 2y)^p}} + x \ln{\frac{x \Gamma}{ e |\epsilon_0|}}}_\text{Typical value of the sum in Eq.~\eqref{eq:forward_scattering_approximation}} \underbrace{ \vphantom{\int_0^x} - x 
\ln{x} - (1 - x)\ln{(1 - x)} - \frac{1 - (1 - 2x)^p}{1 + (1 - 2x)^p} \epsilon_0^2}_\text{Number of degenerate configs. at $x$ (i.e., $s(x)$)}.
\end{equation}
\end{widetext}
Analogous to $s(x)$, $f(x) < 0$ means that every configuration at distance $x$ has an amplitude that vanishes with certainty as $N \rightarrow \infty$.

\begin{figure}[h]
\begin{center}
\includegraphics[width=1.0\columnwidth]{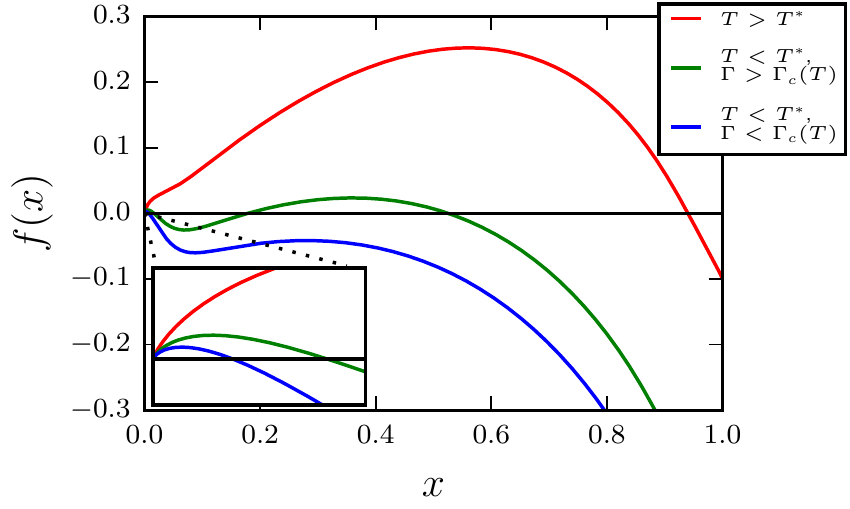}
\caption{The annealed entropy of resonances $f(x)$ (for $p = 20$). $x$ is the fractional Hamming distance relative to an unperturbed state with energy density $\epsilon_0 = -\frac{1}{2T}$. The inset shows $f(x)$ at very small $x$. Red curve: $(T, \Gamma) = (2.27, 0.50)$. Green curve: $(T, \Gamma) = (1.56, 0.50)$. Blue curve: $(T, \Gamma) = (1.39, 0.50)$.}
\label{fig:num_res}
\end{center}
\end{figure}

Fig.~\ref{fig:num_res} shows the three different qualitative behaviors of $f(x)$ as $T$ and $\Gamma$ are varied. 
Each corresponds to one of the cases described above. 
For $T \ll T^*$ (bottom two curves, Fig.~\ref{fig:num_res}), there are resonances at $x < x^*$, belonging to the same cluster (see inset). 
Resonances belonging to different clusters only appear when $\Gamma$ exceeds a critical $\Gamma_c(T)$ (middle curve). In line with the comments above, we should treat the intra-cluster resonances via degenerate perturbation theory before considering larger distances. However, since intra-cluster resonances cannot extend past $x^* \ll O(1)$, this effect only gives subleading corrections to the number of resonances in other clusters.
At $T > T^*$ (top curve), one can no longer define separate clusters and we find resonances throughout configuration space. 
The FSA is certainly not valid in this regime -- it merely confirms the consistency of our results.

It is straightforward to determine $\Gamma_c(T)$ from Eq.~\eqref{eq:num_resonances}: it is defined by where the maximum of $f(x)$ over all $x > x^*(T)$ is 0. The result \citep[Sec.~F]{SuppInfo2017} is the portion of the blue curve below $T^*$ in Fig.~\ref{fig:phase_diagram}. To within $O \left( \frac{1}{p} \right) $ corrections, it is identical to that of the QREM \cite{Baldwin2016}. Since $T^*$ diverges as $p$ increases, we find that the non-ergodic phase of $H_p$ does indeed map continuously onto the MBL phase of $H_{\textrm{QREM}}$.

The fact that eigenstates at low $T$ and small $\Gamma$ are non-ergodic has important consquences for their properties, many of which are commonly associated with MBL. One prominent observable in spin-glass theory is the Edwards-Anderson order parameter $q_{\textrm{EA}} \equiv \frac{1}{N} \sum_i \avg{\sigma_i}^2$, where the average is with respect to the Gibbs distribution. We define an eigenstate variant $q_{\textrm{ES}}(\Psi) \equiv \frac{1}{N} \sum_i \braket{\Psi}{\sigma_i^z | \Psi}^2$. Note that $q_{\textrm{ES}}(\Psi) = q_{\textrm{EA}}$ whenever ETH holds. Heuristically, $q_{\textrm{ES}}(\Psi)$ measures how similar the configurations are for which $\ket{\Psi}$ has significant amplitude. $q_{\textrm{ES}}(\Psi) \sim 1$ means that measuring the $\sigma^z$ configuration within state $\ket{\Psi}$ will consistently give macroscopically similar results. One can then associate a specific magnetization pattern to $\ket{\Psi}$. As shown in \citep[Sec.~G]{SuppInfo2017}, in the non-ergodic phase of $H_p$,
\begin{equation} \label{eq:q_ES_result}
q_{\textrm{ES}} = 1 - \frac{4 \Gamma^2 T^2}{p^2} + \ldots .
\end{equation}
Compare to ergodic eigenstates in the paramagnetic phase, which have $q_{\textrm{ES}} = 0$.

\begin{figure}[h]
\begin{center}
\includegraphics[width=1.0\columnwidth]{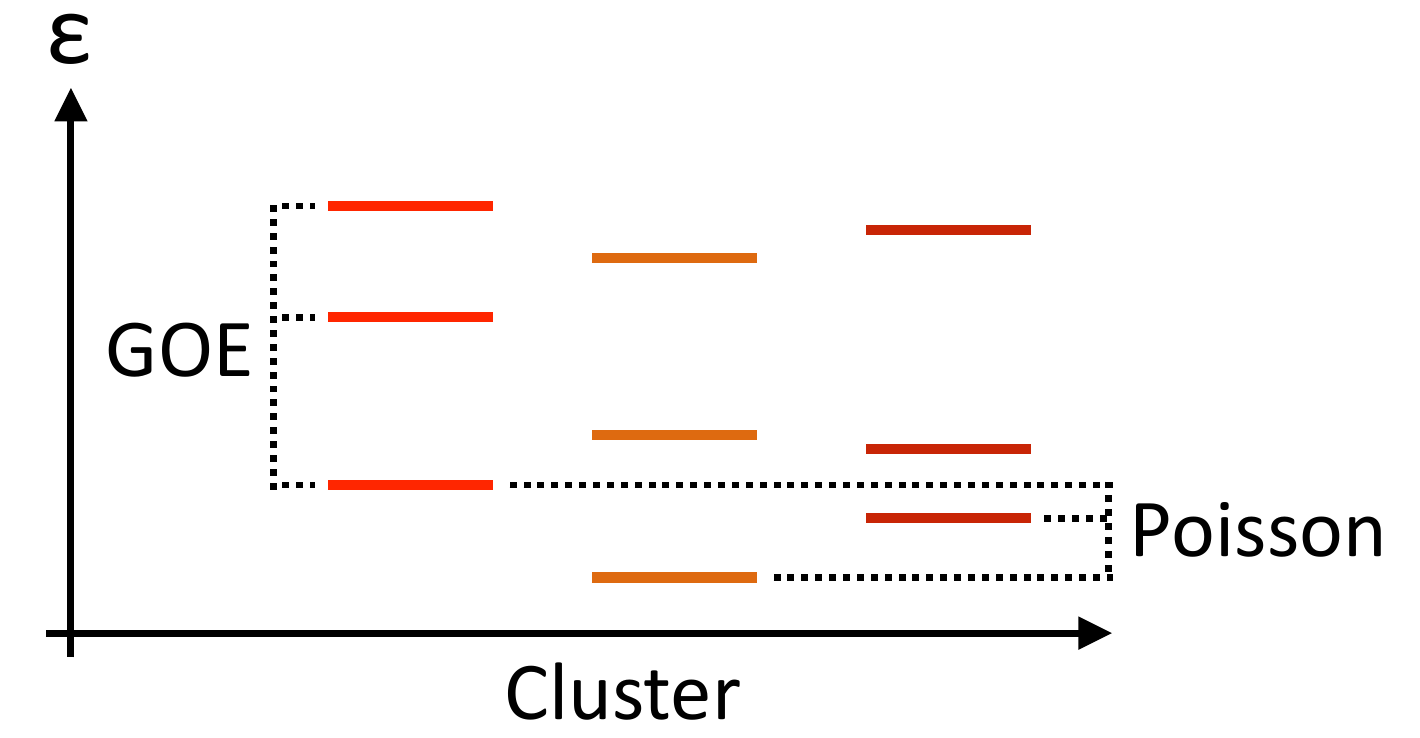}
\caption{A sketch of $H_p$'s eigenvalues in the non-ergodic phase (y-axis), organized by the cluster to which each eigenstate belongs (x-axis).}
\label{fig:clusters}
\end{center}
\end{figure}

The level statistics in the non-ergodic phase is Poisson as well, just as in many-body-localized systems. Regardless of how the eigenstates hybridize within a cluster, they cannot do so over more than the total number of configurations in the cluster, which is $e^{N \exp{(-p \epsilon_0^2)}}$ \citep[Sec.~C]{SuppInfo2017}. These levels strongly repel and have GOE statistics, but the spacing between them scales no smaller than $e^{-N \exp{(-p \epsilon_0^2)}}$. Yet different clusters have independent fluctuations in energy levels, and the number of clusters is at least $e^{N \left( \ln{2} - \epsilon_0^2 - \exp{(-p \epsilon_0^2)} \right) } \gg e^{N \exp{(-p \epsilon_0^2)}}$. The spectra of different clusters interpenetrate, so the level statistics is Poisson. This is sketched in Fig.~\ref{fig:clusters}.

It is important to bear in mind that although we give many analytic results only asymptotically in large $p$, the phenomenology that we have described here applies for \textit{all} $p$, including those most likely to be experimentally realized ($p = 2, 3$) \cite{Strack2011Dicke,Gopalakrishnan2011Frustration}. Existence of the non-ergodic phase relies only on clustering in configuration space, which is known to occur for all $p$ \cite{Gardner1985,Kirkpatrick1987} \footnote{Numerical studies, e.g., exact diagonalization, could provide an important validation, yet for small $p$ one needs very large $N$ to separate the various features. Current state-of-the-art techniques are limited to $N \approx 30$ \cite{Luitz2015,Lerose2015}, which is too small.}. The clustering phenomenon even extends beyond the $p$-spin models, making our results also relevant for, e.g., quantum annealing experiments on combinatorial optimization problems \cite{boixo2014evidence}.

Since the non-ergodic phase looks very similar to a many-body-localized phase, it raises the obvious question of how the two are related. The underlying physics is different: MBL is intrinsically a result of quantum interference, whereas the non-ergodic phase is more a consequence of $O(N)$ energy and entropy barriers. In that respect, it relates more to the \textit{classical} theory of glassiness in mean-field systems.

The relationship to mean-field spin-glass theory, and in particular the replica theory \cite{Mezard1987, Mezard2009}, is potentially very deep. Most prominent is the connection between our $T^*$ and the ``dynamical'' transition temperature $T_d$ \cite{Sompolinksy1981, Kirkpatrick1987, Cugliandolo2004}. Below $T_d$, the Gibbs distribution concentrates around clusters in configuration space (although the equilibrium properties may still be paramagnetic). The classical transition that we identify at $T^*$ also corresponds to clustering in configuration space, even though we have obtained it by independent means. The exact relationship between our calculations and the standard canonical analysis remains to be established, but interestingly, Eq.~\eqref{eq:T_d_behavior} for the asymptotic-in-$p$ behavior of $T^*$ agrees exactly with $T_d$ in the literature \cite{Ferrari2012}. Furthermore, in~\cite{Cugliandolo2004} the authors studied $H_p$ via replica theory and found an entire curve $T_d(\Gamma)$, which in other models was shown to relate to real-time dynamics in the presence of a heat bath \cite{Cugliandolo2001}. That transition lies above the non-ergodic/ETH transition in Fig.~\ref{fig:phase_diagram}; the connection between the two is an interesting open question.

On that note, it was recently argued \cite{chandran2016many} that ergodicity of eigenstates need not imply ergodicity of dynamics. It is possible that the decay time of classical states might diverge in the thermodynamic limit when $T < T^*$, even when the corresponding eigenstates are ergodic. If so, the thermodynamic curve $T_d(\Gamma)$ may describe the quench behavior of $H_p$ rather than eigenstate properties.

And finally, the intra-cluster structure of the non-ergodic eigenstates may be very rich. The ultrametric structure of Parisi's solution \cite{Mezard1984} suggests that a cluster is organized into subclusters, which themselves have subclusters, and so on. The non-ergodic phase may actually be many different phases, with varying degrees of ergodicity-breaking corresponding to how many levels of clusters the eigenstates tunnel through, analogous to the physical picture of replica-symmetry-breaking.

\textit{Acknowledgements} --- We would like to thank L. Cugliandolo, G. Biroli, M. Sellitto, D. Huse, and A. Chandran for helpful discussions. We acknowledge the hospitality of ICTP, where part of this work was completed. C.L.B. also thanks the UW high-performance-computing club for providing necessary computer resources, and the NSF for support through a Graduate Research Fellowship, Grant No. DGE-1256082. C.R.L. acknowledges support from the Sloan Foundation through a Sloan Research Fellowship and the NSF through grant PHY-1520535. The work of A.P. was performed in part at the Aspen Center for Physics, which is supported by National Science Foundation grant PHY-1066293. Note that any opinion, findings, and conclusions or recommendations expressed in this material are those of the authors and do not necessarily reflect the views of the NSF.

\bibliography{p_Spin_Biblio}

\begin{widetext}

\section*{Supplemental Info for ``Clustering of non-ergodic eigenstates in quantum spin glasses''}

\subsection{Thermodynamics in the paramagnetic phase.}

Before studying the eigenstates of the $p$-spin model, we describe the paramagnetic phase of the model's thermodynamics. This analysis comes from Goldschmidt's paper \cite{Goldschmidt1990a}, in which he applies the (imaginary-time) path-integral replica formalism to this model. He also makes the static approximation, i.e., sets the imaginary-time spin autocorrelation to a constant. Yet since the paramagnetic phase is replica-symmetric, we do not include any discussion of replica-symmetry-breaking. 

Starting from Eq.~19 in \cite{Goldschmidt1990a} and setting the RSB parameters $Q = \lambda = 0$, the free energy reduces to
\begin{equation} \label{eq:para_free_energy_start}
f = -\frac{1}{4T} \chi^p + \frac{1}{2T} \chi \nu - T \ln{\int_{-\infty}^{\infty} \frac{\textrm{d}z}{\sqrt{2\pi}} e^{-\frac{1}{2}z^2} 2\cosh{\frac{\sqrt{\Gamma^2 + \nu z^2}}{T}}}.
\end{equation}
$\chi$ is the correlation $\avg{\sigma^z(k) \sigma^z(k')}$ of a single spin between different imaginary times $k$ and $k'$, and $\nu$ is the associated Lagrange multiplier. It is standard to make the static approximation, in which $\avg{\sigma^z(k) \sigma^z(k')}$ is assumed independent of $k - k'$. The equations that minimize $f$ are
\begin{equation} \label{eq:free_energy_equation_1}
\nu = \frac{1}{2} p \chi^{p-1},
\end{equation}
\begin{equation} \label{eq:free_energy_equation_2}
\chi = T \left( \int_{-\infty}^{\infty} \frac{\textrm{d}z}{\sqrt{2\pi}} e^{-\frac{1}{2}z^2} 2\cosh{\frac{\sqrt{\Gamma^2 + \nu z^2}}{T}} \right) ^{-1} \int_{-\infty}^{\infty} \frac{\textrm{d}z}{\sqrt{2\pi}} e^{-\frac{1}{2}z^2} \frac{z^2}{\sqrt{\Gamma^2 + \nu z^2}} 2\sinh{\frac{\sqrt{\Gamma^2 + \nu z^2}}{T}}.
\end{equation}

We expand the integrals in two asymptotic regimes, large $\nu$ and small $\nu$. After straightforward but tedious analysis, we find that
\begin{equation} \label{eq:free_energy_expansion}
f \sim - \frac{1}{4T} \chi^p + \frac{1}{2T} \chi \nu -
\begin{cases}
T \ln{\left( 2 \cosh{\frac{\Gamma}{T}} \right) } + \frac{\nu}{2\Gamma} \tanh{\frac{\Gamma}{T}} , \qquad & \nu \ll \Gamma^2, \Gamma T \\
T \ln{2} + \frac{\nu}{2T} + \frac{\Gamma^2 T}{2 \nu} + \frac{\Gamma^2 T^3}{2 \nu^2} , \qquad & \nu \gg T^2, \Gamma T
\end{cases} .
\end{equation}
The saddle-point equation~\eqref{eq:free_energy_equation_2} (i.e., $\partial_{\nu} f = 0$) then becomes
\begin{equation} \label{eq:free_energy_equation_2_simple}
\chi =
\begin{cases}
\frac{T}{\Gamma} \tanh{\frac{\Gamma}{T}} , \qquad & \nu \ll \Gamma^2, \Gamma T  \\
1 - \frac{\Gamma^2 T^2}{\nu^2} - \frac{2\Gamma^2 T^4}{\nu^3} , \qquad & \nu \gg T^2, \Gamma T
\end{cases}.
\end{equation}
Together with Eq.~\eqref{eq:free_energy_equation_1}, we find two solutions. The quantum paramagnetic solution is $\chi \sim \frac{T}{\Gamma} \tanh{\frac{\Gamma}{T}}$, with free energy
\begin{equation} \label{eq:quantum_free_energy_solution}
f_Q \sim -T \ln{\left( 2\cosh{\frac{\Gamma}{T}} \right) }.
\end{equation}
The classical paramagnetic solution is $\chi \sim 1 - \frac{4 \Gamma^2 T^2}{p^2}$, with free energy 
\begin{equation} \label{eq:classical_free_energy_solution}
f_C \sim -T \ln{2} - \frac{1}{4T} - \frac{\Gamma^2 T}{p} - \frac{2 \Gamma^2 T^3}{p^2}.
\end{equation}
We determine which is the equilibrium solution by setting $f_Q = f_C$. The resulting phase boundary $\Gamma_b(T)$ is always $O(1)$. For $\Gamma < \Gamma_b(T)$, the classical paramagnetic free energy is lower. For $\Gamma > \Gamma_b(T)$, the quantum paramagnetic free energy is lower.

This calculation is only correct when the temperature is low enough for the equilibrium solution to match the asymptotics we assumed. The classical solution requires $\nu_C = \frac{1}{2} p \left( 1 - \frac{4\Gamma^2 T^2}{p^2} \right) ^{p-1} \gg T^2, \Gamma T$. Since it is the equilibrium solution when $\Gamma \lesssim O(1)$, this amounts to $T \ll \sqrt{p}$. Similary, the quantum solution requires $\nu_Q = \frac{1}{2} p \left( \frac{T}{\Gamma} \tanh{\frac{\Gamma}{T}} \right) ^{p-1} \ll \Gamma^2, \Gamma T$, which for $\Gamma \sim O(1)$ amounts to $T \ll \sqrt{\frac{p}{\ln{p}}}$.

For the current paper, we're interested in small $\Gamma$, i.e., the classical paramagnetic phase. So we have
\begin{equation} \label{eq:classical_free_energy}
f = -T \ln{2} - \frac{1}{4T} - \frac{\Gamma^2 T}{p} - \frac{2\Gamma^2 T^3}{p^2} + \cdots , \qquad (T \ll \sqrt{p}) .
\end{equation}
The energy density at temperature $T$ is
\begin{equation} \label{eq:energy_temperature_relation}
\epsilon(T) = -\frac{1}{2T} + \frac{4 \Gamma^2 T^3}{p^2} + \cdots  , \qquad (T \ll \sqrt{p}) .
\end{equation}
The thermodynamic entropy density is
\begin{equation} \label{eq:thermodynamic_entropy}
s_t(T) = \ln{2} - \frac{1}{4T^2} + \frac{\Gamma^2}{p} + \frac{6\Gamma^2 T^2}{p^2} + \cdots , \qquad (T \ll \sqrt{p}) .
\end{equation}
Eq.~\eqref{eq:energy_temperature_relation} is the relationship between energy and temperature used throughout the main text. Note that we only make use of Eq.~\eqref{eq:energy_temperature_relation} for $T \lesssim \sqrt{\frac{p}{\ln{p}}}$, and for those temperatures $\epsilon(T) \sim -\frac{1}{2T}$ to leading order in $\frac{1}{p}$.

\subsection{Distribution of classical energies.}

The classical energy of a configuration $\alpha$ in the $p$-spin model is
\begin{equation} \label{eq:classical_energy}
E_{\alpha} = \sum_{(i_1 \ldots i_p)} J_{i_1 \ldots i_p} \sigma_{i_1}^{z (\alpha)} \cdots \sigma_{i_p}^{z (\alpha)},
\end{equation}
where the sum is over all distinct $p$-tuples of $N$ spins. The couplings $J_{i_1 \ldots i_p}$ are independent Gaussians of mean 0 and variance $\frac{p!}{2N^{p-1}}$. Since $\sigma_{i_1}^{z (\alpha)} \cdots \sigma_{i_p}^{z (\alpha)} = \pm 1$ for each $(i_1 \ldots i_p)$, $E_{\alpha}$ is a sum of Gaussians. Thus it is Gaussian itself, of mean 0 and variance $\binom{N}{p} \frac{p!}{2N^{p-1}} \sim \frac{N}{2}$. In terms of the energy density,
\begin{equation} \label{eq:single_level_distribution}
P_1(\epsilon_{\alpha}) = A_1 \exp{\left( -N \epsilon_{\alpha}^2 \right) },
\end{equation}
where $A_1$ is the normalization.

Now consider the joint distribution for configurations $\alpha$ and $\beta$ that differ in the value of $Nx_{\alpha \beta}$ spins. $E_{\alpha}$ and $E_{\beta}$ are determined by sums over the same couplings. A given term $J_{i_1 \ldots i_p} \sigma_{i_1}^z \cdots \sigma_{i_p}^z$ will contribute the same value for both if an even number of the $Nx_{\alpha \beta}$ differing spins are in the tuple $(i_1 \ldots i_p)$. It will contribute opposite values if an odd number are involved. Let $d_{\textrm{even}}$ denote the number of $p$-tuples that contain an even number of the differing spins, and $d_{\textrm{odd}}$ denote the number that contain an odd number. Then $\frac{E_{\alpha} + E_{\beta}}{2}$ is a sum of $d_{\textrm{even}}$ Gaussians and $\frac{E_{\alpha} - E_{\beta}}{2}$ is a sum of $d_{\textrm{odd}}$ different Gaussians. Note that
\begin{equation} \label{eq:tuple_counting_1}
\begin{aligned}
d_{\textrm{even}} + d_{\textrm{odd}} =& \; \sum_{l = 0: l \textrm{ even}}^p \binom{Nx_{\alpha \beta}}{l} \binom{N(1 - x_{\alpha \beta})}{p - l} + \sum_{l = 1: l \textrm{ odd}}^p \binom{Nx_{\alpha \beta}}{l} \binom{N(1 - x_{\alpha \beta})}{p - l} \\
=& \; \sum_{l=0}^p \binom{Nx_{\alpha \beta}}{l} \binom{N(1 - x_{\alpha \beta})}{p - l} = \binom{N}{p} \sim \frac{N^p}{p!},
\end{aligned}
\end{equation}
\begin{equation} \label{eq:tuple_counting_2}
\begin{aligned}
d_{\textrm{even}} - d_{\textrm{odd}} =& \; \sum_{l = 0: l \textrm{ even}}^p \binom{Nx_{\alpha \beta}}{l} \binom{N(1 - x_{\alpha \beta})}{p - l} - \sum_{l = 1: l \textrm{ odd}}^p \binom{Nx_{\alpha \beta}}{l} \binom{N(1 - x_{\alpha \beta})}{p - l} \\
=& \; \sum_{l=0}^p (-1)^l \binom{Nx_{\alpha \beta}}{l} \binom{N(1 - x_{\alpha \beta})}{p - l} \\
\sim & \; \sum_{l=0}^p (-1)^l \frac{N^p}{l! (p-l)!} x_{\alpha \beta}^l (1 - x_{\alpha \beta})^{p-l} = \frac{N^p}{p!} (1 - 2x_{\alpha \beta})^p.
\end{aligned}
\end{equation}
Thus
\begin{equation} \label{eq:tuple_counting_3}
d_{\textrm{even}} = \frac{N^p}{2p!} (1 + (1 - 2x_{\alpha \beta})^p),
\end{equation}
\begin{equation} \label{eq:tuple_counting_4}
d_{\textrm{odd}} = \frac{N^p}{2p!} (1 - (1 - 2x_{\alpha \beta})^p).
\end{equation}
The variance of $\frac{E_{\alpha} + E_{\beta}}{2}$ is $\frac{N}{4} (1 + (1 - 2x_{\alpha \beta})^p)$, and the variance of $\frac{E_{\alpha} - E_{\beta}}{2}$ is $\frac{N}{4} (1 - (1 - 2x_{\alpha \beta})^p)$. Thus the joint distribution of $\epsilon_{\alpha}$ and $\epsilon_{\beta}$ is
\begin{equation} \label{eq:two_level_distribution}
P_2(\epsilon_{\alpha}, \epsilon_{\beta}) = A_2 \exp{ \left( -\frac{N}{2} \left( \frac{(\epsilon_{\alpha} + \epsilon_{\beta})^2}{1 + (1 - 2x_{\alpha \beta})^p} + \frac{(\epsilon_{\alpha} - \epsilon_{\beta})^2}{1 - (1 - 2x_{\alpha \beta})^p} \right) \right) },
\end{equation}
with $A_2$ the normalization.

We can determine the joint distribution for a general $k$ energies by more formal means. By definition,
\begin{equation} \label{eq:general_joint_distribution_definition}
P_k(\epsilon_1, \ldots , \epsilon_k) = \mathbb{E} \left[ \delta \left( \frac{1}{N} \sum_{(i_1 \cdots i_p)} J_{i_1 \cdots i_p} \sigma_{i_1}^{z(1)} \cdots \sigma_{i_p}^{z(1)} - \epsilon_1 \right) \cdots \delta \left( \frac{1}{N} \sum_{(i_1 \cdots i_p)} J_{i_1 \cdots i_p} \sigma_{i_1}^{z(k)} \cdots \sigma_{i_p}^{z(k)} - \epsilon_k \right) \right] .
\end{equation}
Now write the $\delta$-functions as integrals over Lagrange multipliers $\mu_1, \ldots , \mu_k$:
\begin{equation} \label{eq:k_point_distribution}
\begin{aligned}
P_k(\epsilon_1, \ldots , \epsilon_k) =& \int \mathcal{D} \mu \, e^{-N (\mu_1 \epsilon_1 + \ldots + \mu_k \epsilon_k )} \prod_{(i_1 \cdots i_p)} \mathbb{E} \left[ e^{J_{i_1 \cdots i_p} \left( \mu_1 \sigma_{i_1}^{(1)} \cdots \sigma_{i_p}^{(1)} + \ldots + \mu_k \sigma_{i_1}^{(k)} \cdots \sigma_{i_p}^{(k)} \right) } \right] \\
=& \int \mathcal{D} \mu \, e^{-N (\mu_1 \epsilon_1 + \ldots + \mu_k \epsilon_k ) + \frac{p!}{4N^{p-1}} \sum_{(i_1 \cdots i_p)} \left( \mu_1 \sigma_{i_1}^{(1)} \cdots \sigma_{i_p}^{(1)} + \ldots + \mu_k \sigma_{i_1}^{(k)} \cdots \sigma_{i_p}^{(k)} \right) ^2} \\
=& \; A_k e^{-N \braket{\epsilon}{Q^{-1} | \epsilon}},
\end{aligned}
\end{equation}
where $\ket{\epsilon}$ is the $k$-dimensional vector with entries $\{ \epsilon_1, \ldots , \epsilon_k \}$, and $Q$ is the $k \times k$ matrix with entries
\begin{equation} \label{eq:four_point_matrix}
Q_{ab} = \left( \frac{1}{N} \sum_i \sigma_i^{(a)} \sigma_i^{(b)} \right) ^p = (1 - 2x_{ab})^p.
\end{equation}
This is a closed-form expression for $P_k(\epsilon_1, \ldots , \epsilon_k)$, although inverting the matrix $Q$ is still non-trivial. Note that in the case $k = 2$, we do recover Eq.~\eqref{eq:two_level_distribution}.

\subsection{The annealed distance-resolved entropy. Transition temperatures and cluster properties.}

Here we consider the following question: if a given configuration $\alpha$ has energy density $\epsilon_0$, how many of the other configurations with energy density $\epsilon_0$ are at Hamming distance $Nx$ from $\alpha$? As long as $\epsilon_0$ is above the ground-state energy density, there will be some configurations having $\epsilon_0$ with probability 1. Since the configurations are all statistically equivalent, the fraction of disorder realizations in which $\alpha$ is non-ergodic given it has $\epsilon_0$ is equal to the fraction of all eigenstates at $\epsilon_0$ (over all disorder realizations) that are non-ergodic. Thus we pick some arbitrary reference configuration and condition on it having energy density $\epsilon_0$.

The conditional distribution for a configuration at distance $x$ is
\begin{equation} \label{eq:conditional_distribution}
P_x (\epsilon) = \frac{P_2(\epsilon, \epsilon_0)}{P_1(\epsilon_0)} = \frac{A_2}{A_1} e^{-N \frac{(\epsilon - (1 - 2x)^p \epsilon_0)^2}{1 - (1 - 2x)^{2p}}}.
\end{equation}
The expected number of configurations at $x$ with $\epsilon_0$ is then $e^{Ns(x)}$ with
\begin{equation} \label{eq:annealed_position_entropy}
\begin{aligned}
s(x) \equiv & \; \frac{1}{N} \ln{\mathbb{E} \left[ \# \left( \alpha \textrm{ at } x \textrm{ with } \epsilon_{\alpha} = \epsilon_0 \right) \right] } \\
=& \; \frac{1}{N} \ln{ \binom{N}{Nx} P_x(\epsilon_0)} \\
=& \; -x \ln{x} - (1 - x) \ln{(1 - x)} - \frac{1 - (1 - 2x)^p}{1 + (1 - 2x)^p} \epsilon_0^2.
\end{aligned}
\end{equation}

As $x \rightarrow 0^+$, $s(x) \sim -x \ln{x} > 0$. Also, $s \left( \frac{1}{2} \right) = \ln{2} - \epsilon_0^2 = s_t(T)$, the (annealed) thermodynamic entropy. As long as $\epsilon_0$ lies within the spectrum, $s_t(T) > 0$. Thus $s(x)$ is positive as $x$ approaches both endpoints of $[0, \frac{1}{2}]$, and there is a transition temperature $T^*$ at which $s(x)$ first becomes negative at an intermediate $x$.

We can show that $T^*$ occurs at a scale $\sqrt{\frac{p}{\ln{p}}}$ and $s(x)$ first becomes negative on a scale $\frac{1}{p \sqrt{\ln{p}}}$: set $x = x_r \frac{1}{p \sqrt{\ln{p}}}$ and $\epsilon = \epsilon_r \sqrt{\frac{\ln{p}}{p}}$, with $x_r \sim \Theta(1)$ and $\epsilon_r \sim \Theta(1)$. Then
\begin{equation} \label{eq:position_entropy_expansion}
\begin{aligned}
s(x) =& \; - \frac{x_r}{p \sqrt{\ln{p}}} \ln{\frac{x_r}{p \sqrt{\ln{p}}}} - \left( 1 - \frac{x_r}{p \sqrt{\ln{p}}} \right) \ln{ \left( 1 - \frac{x_r}{p \sqrt{\ln{p}}} \right) } - \frac{1 - \left( 1 - \frac{2x_r}{p \sqrt{\ln{p}}} \right) ^p}{1 + \left( 1 - \frac{2x_r}{p \sqrt{\ln{p}}} \right) ^p} \frac{\ln{p}}{p} \epsilon_r^2 \\
=& \; - \frac{x_r}{p \sqrt{\ln{p}}} \ln{\frac{x_r}{p \sqrt{\ln{p}}}} + \frac{x_r}{p \sqrt{\ln{p}}} - \frac{\ln{p}}{p} \left( \frac{x_r}{\sqrt{\ln{p}}} - \frac{x_r^3}{3 \sqrt{\ln{p}}^3} \right) \epsilon_r^2 + O \left( \frac{1}{p \sqrt{\ln{p}}^3} \right) \\
=& \; \frac{\sqrt{\ln{p}}}{p} \left( \left(1 + \frac{\ln{\ln{p}}}{2 \ln{p}} - \epsilon_r^2 \right) x_r + O \left( \frac{1}{\ln{p}} \right) \right) .
\end{aligned}
\end{equation}
If $\epsilon_r^2 < 1 + \frac{\ln{\ln{p}}}{2 \ln{p}}$, then $s(x)$ increases linearly from 0. Yet if $\epsilon_r^2 > 1 + \frac{\ln{\ln{p}}}{2 \ln{p}}$, then $s(x)$ decreases linearly from 0. This is the transition, which in terms of temperature using Eq.~\eqref{eq:energy_temperature_relation} is
\begin{equation} \label{eq:dynamical_transition_temperature}
T^* = \sqrt{\frac{p}{4 \ln{p}}} \left( 1 + O \left( \frac{\ln{\ln{p}}}{\ln{p}} \right) \right) .
\end{equation}

Below $T^*$, there is an initial region (``cluster'') of configurations at small $x$ and a separate region of configurations at larger $x$. These are separated by a region $(x^*(T), x^{**}(T))$ in which none of the configurations have energy density $\epsilon_0(T)$. Assuming $x \ll \frac{1}{p}$,
\begin{equation} \label{eq:small_distance_position_entropy}
s(x) \sim -x \ln{x} + x - p \epsilon_0^2 x.
\end{equation}
We see that $s(x) > 0$ for $x < x^*(T)$, with
\begin{equation} \label{eq:initial_cluster_size}
x^*(T) = e^{1 - p \epsilon_0^2}.
\end{equation}
This is the length of a typical cluster. As a consistency check, note that $x^*(T)$ is indeed much smaller than $\frac{1}{p}$ so long as $T \ll T^*$.

For $x \gg \frac{1}{p}$,
\begin{equation} \label{eq:large_distance_position_entropy}
s(x) \sim -x \ln{x} - (1 - x) \ln{(1 - x)} - \epsilon_0^2,
\end{equation}
to within exponentially small corrections. $x^{**}(T)$ is the root of this equation, which is $O(1)$ when $\epsilon_0^2 \sim O(1)$. Yet for small $\epsilon_0$,
\begin{equation} \label{eq:next_cluster_distance}
x^{**}(T) \sim \frac{\epsilon_0^2}{\ln{\frac{1}{\epsilon_0^2}}}.
\end{equation}
Note that $x^{**}(T) \gg \frac{1}{p}$ so long as $T \ll T^*$.

Thus $x^{**}(T) - x^*(T) \gg x^*(T)$ when $T \ll T^*$. In words, the separation between clusters is much larger than the length of a cluster. This is not true above $T^*$, where one cannot distinguish different clusters.

Since the expected number of configurations at $x$ is $e^{Ns(x)}$, the total number of configurations within a cluster is dominated by the maximum of $s(x)$, which is
\begin{equation} \label{eq:initial_cluster_num_configs}
\max_{x < x^*(T)} s(x) = e^{-p \epsilon_0^2}.
\end{equation}

\subsection{Rigorous bounds on the presence of clusters}

We can rigorously prove that the classical configurations below $T^*$ are clustered by using moment methods analogous to what was done in Ref.~\cite{Mezard2005} for $k$-SAT problems. Define $W(x)$ to be the number of pairs of configurations separated by distance $x$, both of which have energy density $\epsilon_0$:
\begin{equation} \label{eq:num_pairs_definition}
W(x) \equiv \sum_{\alpha, \beta} \delta_{x_{\alpha \beta}, x} \delta(\epsilon_{\alpha} - \epsilon_0) \delta(\epsilon_{\beta} - \epsilon_0).
\end{equation}
If $W(x)$ is positive for $x$ close to both 0 and $\frac{1}{2}$ but negative for some $x$ in between, then the configurations at $\epsilon_0$ are organized into clusters.

We demonstrate this phenomenon by using two inequalities, one involving the first moment of $W(x)$:
\begin{equation} \label{eq:first_moment_inequality}
\textrm{Pr} \left[ W(x) \geq 1 \right] \leq \mathbb{E} \left[ W(x) \right] ,
\end{equation}
and one involving the second moment:
\begin{equation} \label{eq:second_moment_inequality}
\textrm{Pr} \left[ W(x) \geq 1 \right] \geq \frac{\mathbb{E} \left[ W(x) \right] ^2}{\mathbb{E} \left[ W(x)^2 \right] }.
\end{equation}
The first is Markov's inequality applied to $W(x)$, and the second comes from the fact that the distribution of \emph{non-zero} values for $W(x)$ has non-negative variance. Eq.~\eqref{eq:first_moment_inequality} is useful when $\mathbb{E} \left[ W(x) \right] \rightarrow 0$ in the thermodynamic limit, since then we know that typical samples don't have any pairs separated by $x$. Similarly, Eq.~\eqref{eq:second_moment_inequality} is useful when $\mathbb{E} \left[ W(x)^2 \right] \rightarrow \mathbb{E} \left[ W(x) \right] ^2$, for then there will be at least one pair separated by $x$ with certainty.

The first moment follows from Eq.~\eqref{eq:annealed_position_entropy} in the preceding section. To exponential order,
\begin{equation} \label{eq:first_moment_calculation}
\mathbb{E} \left[ W(x) \right] = 2^N \binom{N}{Nx} P_2(\epsilon_0, \epsilon_0 ) \sim e^{N (\ln{2} - \epsilon_0^2 + s(x) )}.
\end{equation}
For each of the $e^{N(\ln{2} - \epsilon_0^2)}$ configurations having energy density $\epsilon_0$, there are on average $e^{Ns(x)}$ configurations at distance $x$ also with $\epsilon_0$.  We are only interested in $|\epsilon_0| < \sqrt{\ln{2}}$ since we know that the spectrum is contained in this interval. Thus $\mathbb{E} \left[ W(x) \right] \gg 1$ at $x \sim 0$ and $x \sim \frac{1}{2}$. However, for large $p$ and $|\epsilon_0| > \sqrt{\frac{\ln{2}}{2}}$ there is an intermediate region in which $\mathbb{E} \left[ W(x) \right] \ll 1$. If we can use the second moment to show that there \emph{are} pairs at smaller and larger $x$, then we'll have proven the existence of clustering.

From Eq.~\eqref{eq:num_pairs_definition}, we see that $W(x)^2$ involves a sum over sets of four configurations. The orientation between any four is specified by the numbers $x_{mn}, 1 \leq m < n \leq 4$ (with $x_{12} = x_{34} = x$), and we can write
\begin{equation} \label{eq:second_moment_start}
\mathbb{E} \left[ W(x)^2 \right] = \sum_{x_{mn}} \mathcal{N}_4 ( \{ x_{mn} \} ) P_4(\epsilon_0, \epsilon_0, \epsilon_0, \epsilon_0),
\end{equation}
with $P_4$ as given in Eq.~\eqref{eq:k_point_distribution} and $\mathcal{N}_4 ( \{ x_{mn} \} )$ the number of quadruplets separated by the given set of distances. Note that
\begin{equation} \label{eq:num_dumbbells_definition}
\mathcal{N}_4 ( \{ x_{mn} \} ) = \textrm{Tr} \left[ \prod_{m < n} \delta \left( \frac{1}{N} \sum_i \frac{1 + \sigma_i^{z(m)} \sigma_i^{z(n)}}{2} - x_{mn} \right) \right] .
\end{equation}
We expect $\mathcal{N}_4$ to obey a large deviation principle $\mathcal{N}_4 \sim e^{N \Omega_4 ( \{ x_{mn} \} )}$, and note from Eq.~\eqref{eq:k_point_distribution} that $P_4$ does as well. Thus the sum over distances $\{ x_{mn} \}$ in Eq.~\eqref{eq:second_moment_start} is dominated by the saddle-point in $\mathcal{N}_4 P_4$ at $\{ x_{mn}^* \}$ (keeping $x_{12}$ and $x_{34}$ fixed at $x$). Intuitively, the saddle-point in $\mathcal{N}_4$ \emph{by itself} should be at $x_{13}^* = x_{14}^* = x_{23}^* = x_{24}^* = \frac{1}{2}$, and we formally show this below. We then show that $\ln{P_4}$ has vanishing second derivatives with respect to all free distances at $\frac{1}{2}$, meaning that this saddle-point in $\mathcal{N}_4$ is also a saddle-point in $\mathcal{N}_4 P_4$. Yet $\mathcal{N}_4 ( \{ x_{mn}^* \} ) \sim \left( 2^N \binom{N}{Nx} \right) ^2$ and $P_4 \big| _{x_{mn}^*} = P_2 ^2 \big| _x$, thus the saddle-point gives a contribution to $\mathbb{E} \left[ W(x)^2 \right]$ scaling as $\mathbb{E} \left[ W(x) \right] ^2$. If this is the dominant contribution, then the second-moment inequality states that there are configurations separated by $x$ with certainty.

Now we prove these claims. Starting with Eq.~\eqref{eq:num_dumbbells_definition}, introduce Lagrange multipliers $\{ \lambda_{mn} \}$ as done for Eq.~\eqref{eq:k_point_distribution}:
\begin{equation} \label{eq:dumbbell_entropy_formal}
\begin{aligned}
\mathcal{N}_4 =& \int \mathcal{D} \lambda \, e^{-N \sum_{m < n} \lambda_{mn} x_{mn}} \textrm{Tr} \left[ e^{\sum_{m < n} \lambda_{mn} \sum_i \frac{1 + \sigma_i^{z(m)} \sigma_i^{z(n)}}{2}} \right] \\
=& \int \mathcal{D} \lambda \, e^{N \left( - \sum_{m < n} \lambda_{mn} x_{mn} \, + \, \ln{Z_{\textrm{eff}}} \right) },
\end{aligned}
\end{equation}
with
\begin{equation} \label{eq:dumbbell_entropy_effective_partition}
Z_{\textrm{eff}} = \sum_{\sigma^{z(m)}} e^{\sum_{m < n} \lambda_{mn} \frac{1 + \sigma^{z(m)} \sigma^{z(n)}}{2}}.
\end{equation}
Assuming the large-deviation form for $\mathcal{N}_4$, $\max \mathcal{N}_4 \sim \int \mathcal{D}x \, \mathcal{N}_4$, and note that $x_{mn}$ enters into the exponent of Eq.~\eqref{eq:dumbbell_entropy_formal} only linearly, coupled to $\lambda_{mn}$. Thus integrating over $x_{mn}$ gives $\delta \left( \lambda_{mn} \right)$, i.e., we can set $\lambda_{mn} = 0$. Keep in mind that we do not integrate over $x_{12}$ or $x_{34}$ and so do not set $\lambda_{12}$ or $\lambda_{34}$ to $0$. Thus
\begin{equation} \label{eq:Z_eff_factoring}
Z_{\textrm{eff}} \rightarrow \left( \sum_{\sigma^{z(1)}, \sigma^{z(2)}} e^{\lambda_{12} \left( \frac{1 + \sigma^{z(1)} \sigma^{z(2)}}{2} \right) } \right) \left( \sum_{\sigma^{z(3)}, \sigma^{z(4)}} e^{\lambda_{34} \left( \frac{1 + \sigma^{z(3)} \sigma^{z(4)}}{2} \right) } \right)  = 4 \left( 1 + e^{\lambda_{12}} \right) \left( 1 + e^{\lambda_{34}} \right) .
\end{equation}
We integrate over $\lambda_{12}$ and $\lambda_{34}$ by saddle-point, giving $\lambda_{12}^* = \lambda_{34}^* = \ln{\frac{x}{1-x}}$ and
\begin{equation} \label{eq:dumbbell_entropy_max}
\max \left[ \mathcal{N}_4 \right] \sim e^{2N \left( \ln{2} - x \ln{x} - (1 - x)\ln{(1 - x)} \right) } \sim \left( 2^N \binom{N}{Nx} \right) ^2,
\end{equation}
as claimed. Finally, note that if we hadn't integrated over $x_{mn}$, the saddle-point equation for the integral over $\lambda_{mn}$ would have given
\begin{equation} \label{eq:lambda_saddle_point_equation}
x_{mn} = \frac{1}{2} + \sum_{\sigma^{z(u)}} \sigma^{z(m)} \sigma^{z(n)} \frac{e^{\sum_{u < v} \lambda_{uv} \frac{1 + \sigma^{z(u)} \sigma^{z(v)}}{2}}}{Z_{\textrm{eff}}},
\end{equation}
which implicitly defines $\lambda_{mn}^* (x_{mn})$. Since we know that maximizing over $x_{mn}$ fixes $\lambda_{mn}^* = 0$, i.e., $\lambda_{mn}^* (x_{mn}^*) = 0$, we see that $x_{mn}^* = \frac{1}{2}$.

Next, we turn to $P_4$. Note that in Eq.~\eqref{eq:four_point_matrix}, $x_{ab} = \frac{1}{2}$ corresponds to $Q_{ab} = 0$. At the saddle-point $x_{13}^* = x_{14}^* = x_{23}^* = x_{24}^* = \frac{1}{2}$, $Q$ becomes block-diagonal and 
\begin{equation} \label{eq:Q_matrix_factoring}
\braket{\epsilon}{Q^{-1} | \epsilon} \rightarrow \sum_{a,b \in \{ 1, 2 \} } \left( Q^{-1} \right) _{ab} \epsilon_a \epsilon_b + \sum_{a,b \in \{ 3, 4 \} } \left( Q^{-1} \right) _{ab} \epsilon_a \epsilon_b.
\end{equation}
Thus $P_4$ factors: $P_4(\epsilon_1, \epsilon_2, \epsilon_3, \epsilon_4) \rightarrow P_2 (\epsilon_1, \epsilon_2) P_2 (\epsilon_3, \epsilon_4)$. Furthermore, the correction to $Q_{ab}$ from shifting $x_{ab} \rightarrow \frac{1 + \Delta}{2}$ is $O(\Delta^p)$, so we expect that the correction to $\braket{\epsilon}{Q^{-1} | \epsilon}$ is also $O(\Delta^p)$. Indeed, we have confirmed using Mathematica that this is true, uniformly in $x$.

This establishes that $x_{13}^* = x_{14}^* = x_{23}^* = x_{24}^* = \frac{1}{2}$ is a valid saddle-point of $\mathcal{N}_4 P_4$, and thus that $\mathbb{E} \left[ W(x)^2 \right]$ has a contribution scaling as $\mathbb{E} \left[ W(x) \right] ^2$. Since $\frac{1}{N} \ln{P_4} \sim \frac{2}{N} \ln{P_2} + O \left( (1 - 2x_{mn})^p \right)$ for all $m \in \{ 1, 2 \}$, $n \in \{ 3, 4 \}$, we further see that any additional saddle-points to Eq.~\eqref{eq:second_moment_start} must occur at $x_{mn}$ scaling to $0$ with large $p$. Similar analysis to above shows that in the case $x_{13} \rightarrow 0, x_{24} \nrightarrow 0$ (and thus $x_{14} \rightarrow x_{23} \rightarrow x$), $\mathcal{N}_4 P_4 \rightarrow e^{-N(\ln{2} - \epsilon_0^2)} \left( \mathcal{N}_2 P_2 \right) ^2  \ll \left( \mathcal{N}_2 P_2 \right) ^2$. Thus any saddle-points at small $x_{13}$ and not-small $x_{24}$ are sub-leading with respect to that at $\frac{1}{2}$. If we take $x_{13} \rightarrow 0$ \emph{and} $x_{24} \rightarrow 0$, we find that $\mathcal{N}_4 P_4 \rightarrow \mathbb{E} \left[ W(x) \right] $. Assuming that $\mathbb{E} \left[ W(x) \right] \gg 1$, which is certainly true at $x \sim 0$ and $x \sim \frac{1}{2}$ (where $\mathbb{E} \left[ W(x) \right] \sim e^{N (\ln{2} - \epsilon_0^2)}$ and $e^{2N (\ln{2} - \epsilon_0^2)}$, respectively), any saddle-points in this limit are sub-leading as well.

Thus we have demonstrated that $\mathbb{E} \left[ W(x)^2 \right] \sim \mathbb{E} \left[ W(x) \right] ^2$, i.e., that there are pairs of configurations separated by $x$, for $x \sim 0$ and $x \sim \frac{1}{2}$. Together with the observation that there are not pairs separated by an interval of intermediate $x$ for low-enough $\epsilon_0$, this proves the existence of clustering.

\subsection{Replica calculation of the forward-scattering wavefunction.}

From the main text, the forward-scattering expression for an eigenstate $\ket{\Psi}$ of the \underline{quantum} $p$-spin model is
\begin{equation} \label{eq:forward_scattering_general_expression}
\braket{\beta}{\Psi} = \frac{\Gamma}{E_{\alpha} - E_{\beta}} \sum_{\mathcal{P}} \prod_{\gamma_j \in \mathcal{P}} \frac{\Gamma}{E_{\alpha} - E_{\gamma_j}}.
\end{equation}
Here $\alpha$ and $\beta$ are classical configurations of energy $E_{\alpha}$ and $E_{\beta}$, separated by Hamming distance $N x_{\alpha \beta}$. At $\Gamma = 0$, $\ket{\Psi} = \ket{\alpha}$. The sum is over all $(N x_{\alpha \beta})!$ direct ``paths'' (i.e., sequences of spin flips) from configuration $\alpha$ to configuration $\beta$, and $\gamma_j$ is the $j$'th configuration along path $\mathcal{P}$. From now on we'll take $E_{\alpha} \equiv E_0$ and $x_{\alpha \beta} \equiv x$. We'll also assume that the statistics of $\braket{\beta}{\Psi}$, having very heavy tails, aren't significantly affected by cancellations among paths. Thus we make the analytically expedient replacement
\begin{equation} \label{eq:forward_scattering_magnitude_approx}
|\braket{\beta}{\Psi}| \rightarrow \frac{\Gamma}{|E_0 - E_{\beta}|} \sum_{\mathcal{P}} \prod_{\gamma_j \in \mathcal{P}} \frac{\Gamma}{|E_0 - E_{\gamma_j}|}.
\end{equation}

In the main text, we assumed that all paths leading up to $\beta$ have typical amplitudes, and then determined the probability that $|E_0 - E_{\beta}|$ is small enough to make $|\braket{\beta}{\Psi}| > 1$. Here ``typical'' means replacing each $E_{\gamma_j}$ by its mean value, which from Eq.~\eqref{eq:conditional_distribution} is $\left( 1 - \frac{2j}{N} \right) ^p E_0$ (note that configuration $\gamma_j$ by definition differs from $\alpha$ by $j$ spin-flips). We find that
\begin{equation} \label{eq:typical_wavefunction}
|\braket{\beta}{\Psi}|_{\textrm{typ}} = \frac{\Gamma}{|E_0 - E_{\beta}|} (Nx)! \prod_{j=1}^{Nx - 1} \frac{\Gamma}{|E_0| \left( 1 - \left( 1 - \frac{2j}{N} \right) ^p \right) } \equiv \frac{1}{|E_0 - E_{\beta}|} e^{-Nl(x)},
\end{equation}
with
\begin{equation} \label{eq:typical_coupling_exponent}
l(x) \sim -x \ln{\frac{x \Gamma}{e |\epsilon_0|}} + \int_0^x \textrm{d} y \ln{(1 - (1 - 2y)^p)}.
\end{equation}

Here we go beyond the typical amplitudes by making a replica-symmetry-breaking ansatz, to see whether the typical value of the sum in Eq.~\eqref{eq:forward_scattering_magnitude_approx} is approximated by the sum of the typical values. We compute $\frac{1}{N} \mathbb{E} \ln{|\braket{\beta}{\Psi}|} = \frac{1}{Nn} \lim_{n \rightarrow 0} \ln{\mathbb{E} |\braket{\beta}{\Psi}|^n}$ assuming one step of replica-symmetry-breaking. The $n$ replicated paths cluster into $\frac{n}{m}$ independent groups of $m$ identical paths each:
\begin{equation} \label{eq:1RSB_ansatz}
\begin{aligned}
\mathbb{E} |\braket{\beta}{\Psi}|^n =& \; \frac{\Gamma^{Nnx}}{N^{Nnx}} \sum_{\mathcal{P}_1 \cdots \mathcal{P}_n} \mathbb{E} \left( \prod_{\gamma_j \in \mathcal{P}_1} \frac{1}{|\epsilon_0 - \epsilon_{\gamma_j}|} \right) \cdots \left( \prod_{\gamma_j \in \mathcal{P}_n} \frac{1}{|\epsilon_0 - \epsilon_{\gamma_j}|} \right) \\
\rightarrow & \; \frac{\Gamma^{Nnx}}{N^{Nnx}} \sum_{\mathcal{P}_1 \cdots \mathcal{P}_{\frac{n}{m}}} \mathbb{E} \left( \prod_{\gamma_j \in \mathcal{P}_1} \frac{1}{|\epsilon_0 - \epsilon_{\gamma_j}|^m} \right) \cdots \mathbb{E} \left( \prod_{\gamma_j \in \mathcal{P}_{\frac{n}{m}}} \frac{1}{|\epsilon_0 - \epsilon_{\gamma_j}|^m} \right) 
\end{aligned}
\end{equation}
The second line is our replica-symmetry-breaking ansatz. Recall that within this framework, $ 0 \leftarrow n < m \leq 1$. To make further progress, we replace
\begin{equation} \label{eq:independent_weights_approx}
\mathbb{E} \left( \prod_{\gamma_j \in \mathcal{P}} \frac{1}{|\epsilon_0 - \epsilon_{\gamma_j}|^m} \right) \rightarrow \prod_{\gamma_j \in \mathcal{P}} \mathbb{E} \frac{1}{|\epsilon_0 - \epsilon_{\gamma_j}|^m},
\end{equation}
even though the energies along a path are correlated. Although this is not a controlled approximation, it does still let us probe how rare fluctuations might dominate the sum. Denoting $x_j \equiv \frac{j}{N}$,
\begin{equation} \label{eq:moment_evaluation}
\begin{aligned}
\mathbb{E} \frac{1}{|\epsilon_0 - \epsilon_{\gamma_j}|^m} =& \int_{-\infty}^{\infty} \textrm{d} \epsilon \sqrt{\frac{N}{\pi (1 - (1 - 2x_j)^{2p})}} e^{-N \frac{(\epsilon - (1 - 2x_j)^p \epsilon_0)^2}{1 - (1 - 2x_j)^{2p}}} \frac{1}{|\epsilon_0 - \epsilon|^m} \\
=& \sqrt{\frac{N \epsilon_0^2}{\pi (1 - (1 - 2x_j)^{2p})}} |\epsilon_0|^{-m} \int_{-\infty}^{\infty} \textrm{d} z \frac{1}{|1 - z|^m} e^{-N \epsilon_0^2 \frac{(z - (1 - 2x_j)^p )^2}{1 - (1 - 2x_j)^{2p}}} \\
=& \sqrt{\frac{N \epsilon_0^2}{\pi (1 - (1 - 2x_j)^{2p})}} |\epsilon_0|^{-m} \left( \int_{-\infty}^{1} \textrm{d} z \frac{1}{(1 - z)^m} e^{-N \epsilon_0^2 \frac{(z - (1 - 2x_j)^p )^2}{1 - (1 - 2x_j)^{2p}}} \right. \\
& \qquad \qquad \qquad \qquad \qquad \qquad \qquad \left. + \int_{1}^{\infty} \textrm{d} z \frac{1}{(z - 1)^m} e^{-N \epsilon_0^2 \frac{(z - (1 - 2x_j)^p )^2}{1 - (1 - 2x_j)^{2p}}} \right) .
\end{aligned}
\end{equation}
Two things are happening with this expression. The first integral has a saddle-point contribution:
\begin{equation} \label{eq:moment_saddle_point}
\mathbb{E} \frac{1}{|\epsilon_0 - \epsilon_{\gamma_j}|^m} \rightarrow \frac{1}{(1 - (1 - 2x_j)^p)^m |\epsilon_0|^m} + O \left( \frac{1}{N} \right) .
\end{equation}
However, this expression does not diverge as $m \rightarrow 1^-$, yet we know the full expression does. We must be careful that the diverging part of the integral does not get buried in a remainder. For that, we integrate by parts:
\begin{equation} \label{eq:moment_integrate_by_parts}
\begin{aligned}
\int_{-\infty}^{1} \textrm{d} z \frac{1}{(1 - z)^m} e^{-N \epsilon_0^2 \frac{(z - (1 - 2x_j)^p )^2}{1 - (1 - 2x_j)^{2p}}} =& \; \frac{2 N \epsilon_0^2}{1 - (1 - 2x_j)^{2p}} \frac{1}{1 - m} \int_{-\infty}^{1} \textrm{d} z (1 - z)^{1-m} \left( (1 - 2x_j)^p - z \right) e^{-N \frac{(z - (1 - 2x_j)^p )^2}{1 - (1 - 2x_j)^{2p}} \epsilon_0^2} \\
\rightarrow & \; \frac{1}{1 - m} e^{-N \frac{1 - (1 - 2x_j)^p}{1 + (1 - 2x_j)^p} \epsilon_0^2} + O \left( (1 - m)^0 \right) ,
\end{aligned}
\end{equation}
and similarly for the second integral. Putting Eqs.~\eqref{eq:moment_saddle_point} and~\eqref{eq:moment_integrate_by_parts} together, we have that
\begin{equation} \label{eq:moment_asymptotics}
\mathbb{E} \frac{1}{|\epsilon_0 - \epsilon_{\gamma_j}|^m} = \frac{1}{(1 - (1 - 2x_j)^p)^m |\epsilon_0|^m} \left( 1 + \sqrt{\frac{N \epsilon_0^2 b_j}{\pi}} \frac{2}{1 - m} e^{-N \epsilon_0^2 b_j} \right) + O \left( \frac{(1 - m)^0}{N} \right) ,
\end{equation}
where we have defined $b_j \equiv \frac{1 - (1 - 2x_j)^p}{1 + (1 - 2x_j)^p}$.

Eq.~\eqref{eq:moment_asymptotics} goes into Eq.~\eqref{eq:1RSB_ansatz}:
\begin{equation} \label{eq:1RSB_evaluation}
\begin{aligned}
\mathbb{E} |\braket{\beta}{\Psi}|^n \sim & \; \frac{\Gamma^{Nnx}}{N^{Nnx}} (Nx)!^{\frac{n}{m}} \prod_{j=1}^{Nx} \frac{1}{(1 - (1 - 2x_j)^p)^n |\epsilon_0|^n} \left( 1 + \sqrt{\frac{N \epsilon_0^2 b_j}{\pi}} \frac{2}{1 - m} e^{-N \epsilon_0^2 b_j} \right) ^{\frac{n}{m}} \\
\sim & \; \frac{\Gamma^{Nnx}}{(N |\epsilon_0|)^{Nnx}} e^{-n \sum_{j=1}^{Nx} \ln{(1 - (1 - 2x_j)^p)}} \left( \frac{Nx}{e} \right) ^{\frac{Nnx}{m}} e^{\frac{n}{m} \sum_{j=1}^{Nx} \ln{\left( 1 + \sqrt{\frac{N \epsilon_0^2 b_j}{\pi}} \frac{2}{1 - m} e^{-N \epsilon_0^2 b_j} \right) }} \\
\sim & \; \exp{\left( Nn \left( \phi_0(x) + \frac{1}{m} \phi_1(x, m) \right) \right) },
\end{aligned}
\end{equation}
with
\begin{equation} \label{eq:exponent_symmetric}
\phi_0(x) = x \ln{\frac{\Gamma}{N|\epsilon_0|}} - \int_{0}^x \textrm{d} y \ln{(1 - (1 - 2y)^p)},
\end{equation}
\begin{equation} \label{eq:exponent_broken}
\phi_1(x, m) = x \ln{\frac{Nx}{e}} + \frac{1}{N} \sum_{j=1}^{Nx} \ln{\left( 1 + \sqrt{\frac{N \epsilon_0^2 b_j}{\pi}} \frac{2}{1 - m} e^{-N \epsilon_0^2 b_j} \right) }.
\end{equation}

We now \textit{minimize} with respect to $m$ (recall that within the replica formalism we minimize instead of maximize). The minimum exists at a non-trivial value $m_{\textrm{EQ}}$ because $\frac{1}{m} \phi_1(x, m) \rightarrow + \infty$ as $m \rightarrow 0$ and as $m \rightarrow 1$. We need to solve $\frac{1}{m} \partial_m \phi_1 - \frac{1}{m^2} \phi_1 = 0$, which can be written
\begin{equation} \label{eq:m_eq_equation}
N x \ln{\frac{Nx}{e}} + \sum_{j=1}^{Nx} \ln{\left( 1 + \sqrt{\frac{N \epsilon_0^2 b_j}{\pi}} \frac{2}{1 - m_{\textrm{EQ}}} e^{-N \epsilon_0^2 b_j} \right) } = \frac{m_{\textrm{EQ}}}{1 - m_{\textrm{EQ}}} \sum_{j=1}^{Nx} \frac{\sqrt{\frac{N \epsilon_0^2 b_j}{\pi}} \frac{2}{1 - m_{\textrm{EQ}}} e^{-N \epsilon_0^2 b_j}}{1 + \sqrt{\frac{N \epsilon_0^2 b_j}{\pi}} \frac{2}{1 - m_{\textrm{EQ}}} e^{-N \epsilon_0^2 b_j}}.
\end{equation} 
It would be very difficult to fully solve this equation for $m_{\textrm{EQ}}$, but we can pull out how $1 - m_{\textrm{EQ}} \equiv \Delta$ scales with $N$. We'll find that $\Delta \ll 1$. First consider
\begin{equation} \label{eq:replica_tricky_part}
B_j \equiv \sqrt{\frac{N \epsilon_0^2 b_j}{\pi}} \frac{2}{\Delta} e^{-N \epsilon_0^2 b_j}.
\end{equation}
$B_j \gg 1$ (as $\Delta \rightarrow 0$) when $b_j \lesssim \frac{1}{N \epsilon_0^2} \ln{\frac{1}{\Delta}}$. $B_j$ is small otherwise. Assuming $\ln{\frac{1}{\Delta}} \ll N$ (we'll see that this turns out to be true), $b_j \lesssim \frac{1}{N \epsilon_0^2} \ln{\frac{1}{\Delta}}$ corresponds to $j \lesssim \frac{1}{p \epsilon_0^2} \ln{\frac{1}{\Delta}}$. The maximum of $B_j$ is $\sqrt{\frac{2}{\pi e}} \frac{1}{\Delta}$, at $j = \frac{1}{2p \epsilon_0^2}$. Thus we have the bounds
\begin{equation} \label{eq:sum_log_asymptotics}
\frac{C_1}{p \epsilon_0^2} \ln{\frac{1}{\Delta}} \; < \; \sum_{j=1}^{Nx} \ln{\left( 1 + \sqrt{\frac{N \epsilon_0^2 b_j}{\pi}} \frac{2}{\Delta} e^{-N \epsilon_0^2 b_j} \right) } < \; \frac{C_2}{p \epsilon_0^2} \left( \ln{\frac{1}{\Delta}} \right) ^2, 
\end{equation}
for suitable choices of the constants $C_1$ and $C_2$. Since $\ln{\frac{1}{\Delta}} \ll N$, we see that this sum is negligible compared to $Nx \ln{\frac{Nx}{e}}$ in Eq.~\eqref{eq:m_eq_equation}. Similarly, the RHS of Eq.~\eqref{eq:m_eq_equation} is bounded by
\begin{equation} \label{eq:sum_fraction_asymptotics}
\frac{D_1}{p \epsilon_0^2} \frac{1}{\Delta} \ln{\frac{1}{\Delta}} \; < \; \frac{1 - \Delta}{\Delta} \sum_{j=1}^{Nx} \frac{\sqrt{\frac{N \epsilon_0^2 b_j}{\pi}} \frac{2}{\Delta} e^{-N \epsilon_0^2 b_j}}{1 + \sqrt{\frac{N \epsilon_0^2 b_j}{\pi}} \frac{2}{\Delta} e^{-N \epsilon_0^2 b_j}} < \; \frac{D_2}{p \epsilon_0^2} \frac{1}{\Delta} \ln{\frac{1}{\Delta}} ,
\end{equation}
for suitable choices of the constants $D_1$ and $D_2$. Since $\Delta$ is chosen so that this sum is essentially $Nx \ln{\frac{Nx}{e}}$,
\begin{equation} \label{eq:delta_bounds}
\frac{D_1}{p \epsilon_0^2} \frac{1}{\Delta} \ln{\frac{1}{\Delta}} < Nx \ln{\frac{Nx}{e}} < \frac{D_2}{p \epsilon_0^2} \frac{1}{\Delta} \ln{\frac{1}{\Delta}}.
\end{equation}
We see that $\Delta \sim \frac{1}{Nx}$, so that
\begin{equation} \label{eq:m_eq_solution}
m_{\textrm{EQ}} = 1 - \frac{\delta (Nx, p \epsilon_0^2)}{Nx},
\end{equation}
with $\delta (\cdot, \cdot)$ an $O(1)$ function. Although $\delta (Nx, p \epsilon_0^2)$ does vary with $N$, it stays $O(1)$ as $N \rightarrow \infty$.

Thus
\begin{equation} \label{eq:phi_1_evaluation}
\frac{1}{m_{\textrm{EQ}}} \phi_1(x, m_{\textrm{EQ}}) = x \ln{\frac{Nx}{e}} + O \left( \frac{(\ln{N})^2}{N} \right) .
\end{equation}
Our final result for the average log-amplitude, which is to be compared with the ``typical'' estimate in Eqs.~\eqref{eq:typical_wavefunction} and~\eqref{eq:typical_coupling_exponent}, is
\begin{equation} \label{eq:1RSB_amplitude_result}
\begin{aligned}
\frac{1}{N} \mathbb{E} \ln{|\braket{\beta}{\Psi}|} =& \; \frac{1}{Nn} \lim_{n \rightarrow 0} \ln{\mathbb{E} |\braket{\beta}{\Psi}|^n} = \phi_0(x) + \frac{1}{m_{\textrm{EQ}}} \phi_1(x, m_{\textrm{EQ}}) \\
=& \; x \ln{\frac{x \Gamma}{e|\epsilon_0|}} - \int_{0}^x \textrm{d} y \ln{(1 - (1 - 2y)^p)} + O \left( \frac{(\ln{N})^2}{N} \right) .
\end{aligned}
\end{equation}
We see that the corrections to Eq.~\eqref{eq:typical_coupling_exponent} vanish in the thermodynamic limit, even when allowing for one step of replica-symmetry-breaking. This justifies our use of the ``typical'' estimate in the main text.

\subsection{Non-ergodic phase boundary.}

As justified in the previous section, the amplitude of eigenstate $\ket{\Psi}$ on configuration $\ket{\beta}$ is
\begin{equation} \label{eq:wavefunction_amplitude}
|\braket{\beta}{\Psi}| = \frac{1}{|E_0 - E_{\beta}|} e^{-Nl(x)},
\end{equation}
with $l(x)$ given by Eq.~\eqref{eq:typical_coupling_exponent}. Here $E_0$ is the energy of the eigenstate, related to temperature by Eq.~\eqref{eq:energy_temperature_relation}, and $Nx$ is the Hamming distance from $\beta$ to the unperturbed ($\Gamma = 0$) configuration. Assuming $e^{-Nl(x)} \ll 1$ (which it is for all $T$ and $\Gamma$ of interest), $|E_0 - E_{\beta}|$ must be exponentially small for $\beta$ to be resonant. The distribution of $E_{\beta}$ is essentially uniform on such small scales, so a fraction $e^{-Nl(x)}$ of the configurations at distance $x$ with energy density $\epsilon_0$ resonate. There are $e^{Ns(x)}$ such configurations, with $s(x)$ given by Eq.~\eqref{eq:annealed_position_entropy}, so the average number of resonant configurations at distance $x$ is $e^{Nf(x)}$, with
\begin{equation} \label{eq:num_resonances_exponent}
f(x) = s(x) - l(x) = x \ln{\frac{\Gamma}{e|\epsilon_0|}} - (1 - x) \ln{(1 - x)} - \int_{0}^x \textrm{d} y \ln{(1 - (1 - 2y)^p)} - \frac{1 - (1 - 2x)^p}{1 + (1 - 2x)^p} \epsilon_0^2.
\end{equation}

If $f(x) < 0$, then the probability of having \textit{any} resonance at distance $x$ is 0 in the thermodynamic limit. If $f(x) > 0$, then there are resonances at distance $x$ and the eigenstate $\ket{\Psi}$ is hybridized among them. The forward-scattering expression Eq.~\eqref{eq:forward_scattering_general_expression} is then invalid, yet one could in principle write down an analogous expression in terms of ``paths'' among hybridized states. These states would then hybridize if further resonances are found, etc.

For now, consider $p$ large and $x$ independent of $p$. Then $(1 - 2x)^p \rightarrow 0$ and
\begin{equation} \label{eq:num_res_integral_large_limit}
\begin{aligned}
\int_0^x \textrm{d} y \ln{(1 - (1 - 2y)^p)} =& \; \frac{1}{p} \int_0^{px} \textrm{d} z \ln{\left( 1 - \left( 1 - \frac{2z}{p} \right) ^p \right) } \\
=& \; \frac{1}{p} \int_0^{\infty} \textrm{d} z \ln{\left( 1 - e^{-2z} \right) } + O \left( \frac{1}{p^2} \right) = - \frac{\pi^2}{12 p} + O \left( \frac{1}{p^2} \right) .
\end{aligned}
\end{equation}
Thus
\begin{equation} \label{eq:num_res_expression_large_limit}
f(x) = x \ln{\frac{\Gamma}{e |\epsilon_0|}} - (1 - x) \ln{(1 - x)} - \epsilon_0^2 + \frac{\pi^2}{12p} + O \left( \frac{1}{p^2} \right) .
\end{equation}
Note that this expression is valid only for $x \sim O(1)$ with respect to $\frac{1}{p}$. We recover the expression previously obtained for the REM as $p \rightarrow \infty$ \cite{Baldwin2016}. As done in Ref.~\cite{Baldwin2016}, we identify any eigenstate with $f(x) > 0$ for some $x \sim O(1)$ as ergodic. We consider the case of $x \ll 1$ separately. The maximum of Eq.~\eqref{eq:num_res_expression_large_limit} is at $x = 1 - \frac{|\epsilon_0|}{\Gamma}$, with a value of
\begin{equation} \label{eq:num_res_max_large_limit}
f_{\textrm{max}} = \ln{\frac{\Gamma}{|\epsilon_0|}} - 1 + \frac{|\epsilon_0|}{\Gamma} - \epsilon_0^2 + \frac{\pi^2}{12p} + O \left( \frac{1}{p^2} \right) .
\end{equation}
$\Gamma_c(\epsilon_0)$ (expressed as $\Gamma_c(T)$ in the main text) is defined by $f_{\textrm{max}} = 0$. For small $\epsilon_0$,
\begin{equation} \label{eq:phase_boundary}
\Gamma_c(\epsilon_0) = |\epsilon_0| \left( 1 + \sqrt{2 \epsilon_0^2 - \frac{\pi^2}{6p}} + O (\epsilon_0^2) \right) \quad \Leftrightarrow \quad \Gamma_c(T) = \frac{1}{2T} \left( 1 + \sqrt{\frac{2}{T^2} - \frac{\pi^2}{6p}} + O \left( \frac{1}{T^2} \right) \right) .
\end{equation}
The corrections to the REM phase boundary are only $O \left( \frac{1}{p} \right) $, yet we see that for any fixed $p$, this expression breaks down at $|\epsilon_0| < \frac{\pi}{\sqrt{12p}}$. In fact, problems emerge outside of this range as well, due to the behavior of $f(x)$ at small $x$.

For $x \ll \frac{1}{p}$,
\begin{equation} \label{eq:num_res_expression_small_limit}
f(x) = x \ln{\frac{\Gamma}{2p|\epsilon_0|x}} + x - p \epsilon_0^2 x + O (p^2 x^2) .
\end{equation}
This expression is positive for $x < x_{\textrm{res}}^*$, where
\begin{equation} \label{eq:initial_resonances_length}
x_{\textrm{res}}^* = \frac{\Gamma}{2p |\epsilon_0|} e^{1 - p \epsilon_0^2}.
\end{equation}
Compare to $x^*$ in Eq.~\eqref{eq:initial_cluster_size}. Thus there are always resonances at small enough $x$, which lie in the local cluster of configurations at $x < x^*$. Suppose for now that $\epsilon_0 \sim O(1)$, so that $x_{\textrm{res}}^*$ and $x^*$ are exponentially small in $p$. We have no way of describing the wavefunction amplitude on these resonant configurations, and restarting the perturbation series around them will undoubtedly put additional configurations into resonance. But regardless of how the wavefunction hybridizes, there cannot be any resonances at $x \in (x^*, x^{**})$, simply because none of those configurations have energy density $\epsilon_0$. Thus the entropy of resonances $f(x)$ at $x > x^{**}$ is only corrected by an exponentially small amount if we start the forward-scattering at \textit{any} short-distance resonance, even under the worst-case assumption that the resonances extend to $x^*$. And furthermore, we show in the section below that the properties of non-ergodic eigenstates are controlled by the amplitude at distances much larger than $x^*$, so the short-distance resonances do not affect observables either. All this is to say that, at least for $\epsilon_0 \sim O(1)$, none of our conclusions above or in the main text are affected by the short-distance resonances that we're unable to correct for.

However, this reasoning relies on the separation between clusters being much larger than the length of a cluster, which breaks down when $|\epsilon_0| \lesssim \sqrt{\frac{\ln{p}}{p}}$. Thus our description of the non-ergodic eigenstates, in particular Eq.~\eqref{eq:phase_boundary} and the section below, only applies to those with $|\epsilon_0| \gg \sqrt{\frac{\ln{p}}{p}}$.

\subsection{Observables in the non-ergodic phase.}

The non-ergodic eigenstates have a highly-correlated ``core'' that extends no farther than $x^*$, and also a tail in which the wavefunction amplitudes are typical. Any amplitude at distance $x > x^*$ must lie between $e^{-Nl(x + x^*)}$ and $e^{-Nl(x - x^*)}$ (see Eq.~\eqref{eq:typical_wavefunction}), since these are the extreme locations within the core from which one could forward-scatter. Assuming $x \gg x^*$, the leading behavior of the exponent is simply $-l(x)$. Then the total weight in the tail of the wavefunction is
\begin{equation} \label{eq:wavefunction_tail_weight_total}
\int_{x^*} \textrm{d}x \binom{N}{Nx} \left( e^{-N l(x)} \right) ^2 = \int_{x^*}^1 \textrm{d}x \exp{\left( N \left( x \ln{\frac{x \Gamma^2}{e^2 \epsilon_0^2}} - (1 - x) \ln{(1 - x)} - 2 \int_0^x \textrm{d}y \ln{(1 - (1 - 2y)^p)} \right) \right) }.
\end{equation}
The saddle-point is at $x_w \sim \frac{\Gamma^2}{4p^2 \epsilon_0^2}$. For now, assume that $\Gamma$ is not exponentially small in $p$, so that $x_w$ is indeed much larger than $x^*$. The total weight in the tail scales as
\begin{equation} \label{eq:wavefunction_tail_weight_saddle}
\exp{\left( N \frac{\Gamma^2}{4p^2 \epsilon_0^2} \right) }.
\end{equation}
Even if we assume every configuration in the core has amplitude 1 (i.e., as much weight as the unperturbed state), the core cannot contribute more weight than $\binom{N}{Nx^*} \sim \exp{\left( N p \epsilon_0^2 e^{1 - p \epsilon_0^2} \right) } \ll \exp{\left( N \frac{\Gamma^2}{4p^2 \epsilon_0^2} \right) }$. Thus the tail of the wavefunction contains all the weight. This allows us to compute observables.

The inverse participation ratio (IPR) of an eigenstate is defined as
\begin{equation} \label{eq:IPR_definition}
Y_2(\Psi) = \sum_{\alpha} \frac{|\braket{\alpha}{\Psi}|^4}{\left( \sum_{\beta} |\braket{\beta}{\Psi}|^2 \right) ^2}.
\end{equation}
The IPR quantifies how many basis states $\ket{\alpha}$ the wavefunction $\ket{\Psi}$ is spread over. In our case, the Hilbert space has dimension $2^N$. A wavefunction with $Y_2 = 2^{-N}$ has equal weight on all basis states, whereas a wavefunction with $Y_2 = 1$ only has weight on a single basis state. As before, we find that $\sum_{\alpha} |\braket{\alpha}{\Psi}|^4$ is dominated by states in the tail:
\begin{equation} \label{eq:IPR_calculation}
\begin{aligned}
\int_{x^*}^1 \textrm{d}x \binom{N}{Nx} \left( e^{-N l(x)} \right) ^4 =&\; \int_{x^*}^1 \textrm{d}x \exp{\left( N \left( x \ln{\frac{x^3 \Gamma^4}{e^4 \epsilon_0^4}} - (1 - x) \ln{(1 - x)} - 4 \int_0^x \textrm{d}y \ln{(1 - (1 - 2y)^p)} \right) \right) } \\
\sim & \; \exp{\left( N \frac{\Gamma^4}{16p^4 \epsilon_0^4} \right) }
\end{aligned}
\end{equation}
Thus
\begin{equation} \label{eq:IPR_value}
Y_2 \sim \exp{\left( -N \frac{\Gamma^2}{2p^2 \epsilon_0^2} \left( 1 - \frac{\Gamma^2}{8p^2 \epsilon_0^2} \right) \right) }.
\end{equation}
Compare $\ln{Y_2^{-1}}$ with the thermodynamic entropy (Eq.~\eqref{eq:thermodynamic_entropy}). We see that although the non-ergodic wavefunction spreads over exponentially many configurations, it still covers an exponentially small fraction of the total number at $\epsilon_0$.

We next consider the quantity $q_{\textrm{ES}}(\Psi)$ defined as
\begin{equation} \label{eq:q_ES_definition}
q_{\textrm{ES}}(\Psi) = \frac{1}{N} \sum_i \left( \frac{\braket{\Psi}{\hat{\sigma}_i^z | \Psi}}{\braket{\Psi}{\Psi}} \right) ^2.
\end{equation}
This is an eigenstate analogue to the Edwards-Anderson order parameter $q_{\textrm{EA}} \equiv \frac{1}{N} \sum_i \avg{\sigma_i^z}^2$. Whereas the average in $q_{\textrm{EA}}$ is over the Gibbs distribution, in Eq.~\eqref{eq:q_ES_definition} we average over the distribution defined by the eigenstate $\ket{\Psi}$. $q_{\textrm{ES}}$ gives the overlap between two independent measurements of the $\sigma^z$ configuration within $\ket{\Psi}$. To see this, define $c_{\alpha} \equiv \frac{\braket{\alpha}{\Psi}}{\sqrt{\braket{\Psi}{\Psi}}}$, where $\ket{\alpha}$ is a $\sigma^z$ basis state. Then
\begin{equation} \label{eq:q_ES_rewrite}
q_{\textrm{ES}}(\Psi) = \frac{1}{N} \sum_i \left( \sum_{\alpha} \sigma_i^z(\alpha) |c_{\alpha}|^2 \right) ^2 = \sum_{\alpha, \beta} |c_{\alpha}|^2 |c_{\beta}|^2 \left( \frac{1}{N} \sum_i \sigma_i^z(\alpha) \sigma_i^z(\beta) \right) .
\end{equation}
The quantity $\frac{1}{N} \sum_i \sigma_i^z(\alpha) \sigma_i^z(\beta) \equiv Q_{\alpha \beta}$ is the spin overlap between configurations $\alpha$ and $\beta$, exactly as in mean-field spin glass theory.

To evaluate $q_{\textrm{ES}}$ in a non-ergodic eigenstate, first note that $Q_{\alpha \beta} = 1 - 2x_{\alpha \beta}$, with $x_{\alpha \beta}$ the fractional Hamming distance between $\alpha$ and $\beta$. Then
\begin{equation} \label{eq:q_ES_calculation}
q_{\textrm{ES}} = e^{-N \frac{\Gamma^2}{2p^2 \epsilon_0^2}} \int \textrm{d}x \binom{N}{Nx} \int \textrm{d}y \int_{|x - y|}^{x + y} \textrm{d}z \binom{Nx}{N \frac{x - y + z}{2}} \binom{N(1 - x)}{N \frac{-x + y + z}{2}} e^{-2Nl(x)} e^{-2Nl(y)} (1 - 2z) . 
\end{equation}
Here $x$ is the distance to $\alpha$ from the origin, $y$ is the distance to $\beta$ from the origin, and $z$ is the distance from $\alpha$ to $\beta$. Note that $z$ must be between $|x - y|$ and $x + y$, and the number of $\beta$ that satisfy this geometry for each $\alpha$ is $\binom{Nx}{N \frac{x - y + z}{2}} \binom{N(1 - x)}{N \frac{-x + y + z}{2}}$. The saddle-point in $z$ is at $z^* = x + y - 2xy$, and $\binom{Nx}{N \frac{x - y + z^*}{2}} \binom{N(1 - x)}{N \frac{-x + y + z^*}{2}} = \binom{N}{Ny}$ (to leading order). We then have separate saddle-point integrals over $x$ and $y$, both of which are dominated by $x_w = \frac{\Gamma^2}{4p^2 \epsilon_0^2}$. The end result is that
\begin{equation} \label{eq:q_ES_result}
q_{\textrm{ES}} = 1 - \frac{\Gamma^2}{p^2 \epsilon_0^2} + \cdots .
\end{equation}

As mentioned above, these estimates do implicitly assume that $\Gamma \gg e^{-p \epsilon_0^2}$. If $\Gamma$ is exponentially small in $p$, then the bulk of the wavefunction amplitude is in the core and these estimates do not apply. However, the wavefunctions are still non-ergodic, and the qualitative results in this section hold: $\ln{Y_2^{-1}}$ is much smaller than the thermodynamic entropy, and $q_{\textrm{ES}}$ is close to 1.

\end{widetext}

\end{document}